\documentclass[11pt]{amsart}

\usepackage{macros}

\usepackage[utf8]{inputenc}
\usepackage[T1]{fontenc}
\usepackage[backend=biber,
            isbn=false,
            doi=false,
            maxbibnames=5,
            style=alphabetic,
            citestyle=alphabetic]{biblatex}
\usepackage{url}
\renewbibmacro{in:}{}
\bibliography{references.bib}

\usepackage{xcolor}
\definecolor{e-mail}{rgb}{0,.40,.80}
\definecolor{reference}{rgb}{.40,.60,.2}
\definecolor{citation}{rgb}{0,.25,.5}

\usepackage[colorlinks=true,
            linkcolor=reference,
            citecolor=citation,
            urlcolor=e-mail]{hyperref}
\usepackage{cleveref}

\usepackage[top=1in, bottom=1.25in, left=1.3in, right=1.3in]{geometry}
\numberwithin{equation}{section}

\def\d{{\rm d}}

\begin{document}


\title{Quantization of topological-holomorphic field theories: local aspects.}

\author{Owen Gwilliam}
\address{Department of Mathematics and Statistics \\
Lederle Graduate Research Tower, 1623D \\
University of Massachusetts Amherst \\
710 N. Pleasant Street}
\email{gwilliam@math.umass.edu}

\author{Eugene Rabinovich}
\address{Department of Mathematics\\
University of California, Berkeley\\
970 Evans Hall \#3840\\
Berkeley, CA 94720}
\email{erabin@math.berkeley.edu}

\author{Brian R. Williams}
\address{School of Mathematics\\
University of Edinburgh \\ 
Edinburgh \\ 
UK}
\email{brian.williams@ed.ac.uk}

\begin{abstract}
In both mathematics and physics, topological field theories and holomorphic field theories appear naturally,
but there are interesting theories that are hybrids---looking topological in some directions and holomorphic in others---such as twists of supersymmetric field theories or Costello's 4-dimensional Chern--Simons theory.
In this paper we construct perturbative, one-loop quantizations rigorously on the model manifold $\RR^m \times \CC^n$, and find a remarkable vanishing result about anomalies:
the one-loop obstruction to quantization on $\RR^m\times \CC^n$ vanishes when $m\geq 1$.
A concrete consequence of our results is the existence of exact and finite quantizations at one-loop for twists of pure $\cN =2$ four-dimensional supersymmetric Yang--Mills theory.
\end{abstract}

\maketitle
\thispagestyle{empty}
\setcounter{tocdepth}{1}
\tableofcontents

\spacing{1.1}

\section{Introduction}

Topological field theories and holomorphic field theories have each had a substantial impact in both physics and mathematics,
so it is natural to consider theories that are hybrids of the two.
To give a sense of what we mean, imagine a theory living on a product manifold $M \times X$ where the theory depends on the smooth structure of $M$ (but not on any further geometric structure on $M$) and on a complex structure on $X$.
Such theories include twists of $N=2$ four-dimensional supersymmetric Yang-Mills theories \autocite{Kapustin}, 
the four-dimensional Chern--Simons theory of Costello \autocite{CostelloYangian}, Chern--Simons/matter theories with chiral boundary conditions \autocite{CDFK, ACMV, Costello:2020ndc}, and topological-holomorphic AKSZ field theories, among others. 
In a companion paper we will examine global aspects of such topological-holomorphic field theories (THFTs), where the product structure of the manifold $M \times X$ can be generalized to that of a transversely holomorphic foliation. 
In this paper we focus on quantizing these theories  on $\RR^m \times \CC^n$,
where quantization means here rigorous renormalization in conjunction with the Batalin-Vilkovisky (BV) formalism, along the lines of \autocite{CosBook}.
What we show is that thanks to a convenient gauge-fixing condition, the analytic aspects of THFTs are well-behaved,
specifically that one-loop diagrams have no ultraviolet (or short distance) divergences and that the one-loop anomaly vanishes whenever $m > 0$ (i.e., whenever there is at least one topological direction).

A careful statement of our key results is, however, rather technical, in part because they apply at a high level of generality, so we defer stating them until all the definitions and hypotheses are in place.
Instead, in this introduction, we describe corollaries of our key results that may help orient the reader more familiar with the specific theories to which our results apply than with the general theoretical framework within which we prove our results.
A special case of these techniques yields an exact, one-loop quantization of 3-dimensional Chern--Simons theory~\autocite{GWcs};
this paper offers a natural level of generality at which to extend the bounds and tricks from that paper.

To orient the reader, we remark that our gauge fixing condition is analogous to the Lorenz gauge used in electromagnetism and other gauge theories.
Recall that for a connection one-form on a Riemannian or Lorentzian manifold, the Lorenz gauge condition is $\d^\star A = 0$, where $\star$ is the Hodge star operator. 
For a THFT on the product manifold $\RR^m \times \CC^m$, the gauge fields we consider are only {\em partial} connections in the sense that they do not have all the components of a one-form:
they admit an expansion of the form 
\[
A = \sum_i A_{x_i} \d x_i + \sum_j A_{\zbar_j} \d \zbar_j 
\]
where $\{x_i\}$ is a coordinate on $\RR^m$ and $\{z_i\}$ is a holomorphic coordinate on $\CC^n$.
(By comparison, a $\dbar$-connection has only a $(0,1)$-component.)
For such one-forms, our gauge fixing condition becomes
\[
\sum_i \frac{\partial A_{x_i}}{\partial x_i} + \sum_j \frac{\partial A_{\zbar_j}}{\partial z_j} = 0 .
\]
This condition is the restriction of the Lorenz gauge to partial connections where we use the product of the flat metric on $\RR^m$ with the standard Hermitian metric on $\CC^n$. 
We will provide a full BV extension of this condition to an arbitrary THFT on~$\RR^m \times \CC^n$. 

\subsection*{Mixed Chern--Simons theories}

Here, we describe THFTs in dimensions 4 and 5 that interpolate between 
topological Chern--Simons theory on oriented 3-manifolds and holomorphic Chern--Simons theory on Calabi-Yau manifolds of complex dimension 3 (i.e., real dimension 6).
We describe these theories first in local coordinates on $\RR^m \times \CC^n$ where~$m+n = 3$ and then globally on manifolds of the form $M^m\times X^n$.

Fix a complex Lie algebra $\fg$ equipped with an invariant nondegenerate symmetric bilinear form $\kappa$ (e.g., the Killing form on a semisimple Lie algebra like $\sl_n$).
We work in the framework of perturbative field theory, so it is not necessary to specify a Lie group $G$ whose Lie algebra is~$\fg$.

The 4-dimensional Chern--Simons theory for $\fg$ on $\RR^2 \times \CC$ looks like topological Chern--Simons theory where the third coordinate among $(x,y,z)$ is promoted to a complex coordinate.
In explicit terms, the gauge field is a $\fg$-valued 1-form of the following type:
\[
A = A_x \d x + A_y \d y + A_{\zbar} \d \zbar
\]
where each coefficient $A_\bullet$ is a smooth $\fg$-valued function.
Note that a potential term of the form $A_z \d z$ is not present.
The action functional is
\[
S(A) = \int_{\RR^2 \times \CC} \d z \wedge \left( \frac{1}{2}\kappa\left(A, \d_{\RR^2} A + \dbar A \right) + \frac{1}{3}\kappa\left( A, [A,A]\right) \right),
\]
where
\begin{align*}
\d_{\RR^2} A  & = \d x\frac{\partial A}{\partial x}  + \d y\frac{\partial A}{\partial y}  \\
\dbar A & = \d \zbar\frac{\partial A}{\partial \zbar} .
\end{align*}
The equation of motion is then that
\[
\d_{\RR^2} A + \dbar A + \frac{1}{2} [A,A] = 0 \in \Omega^2 (\RR^2 \times \CC) \otimes \fg ,
\]
which picks out a connection that is flat along the $x,y$ directions of $\RR^2$ and holomorphic along the $z$ direction of~$\CC$.
To see this claim explicitly, we expand the one-form $A$ in coordinates and read off the component-wise equations of motion:
\begin{align*}
\partial_x A_y - \partial_y A_x + [A_x, A_y] & = 0 \\
\partial_x A_\zbar - \partial_{\zbar} A_x + [A_x, A_\zbar] & = 0 \\
\partial_y A_\zbar - \partial_\zbar A_y + [A_y, A_\zbar] & = 0 .
\end{align*}
The meaning of the first equation is that, for fixed $z$, the connection $A_x(x,y,z)\d x+A_y(x,y,z)\d y$ on $\RR^2$ is flat.
The meaning of the second and third equations is that, up to a gauge transformation with parameter $A_\zbar$, this connection is holomorphic in its $z$-coordinate.
To quantize this theory, one must choose a gauge-fixing condition. 
At the level of one-forms, our Lorenz-type gauge reads as
\[
\partial_x A_x + \partial_y A_y + 4 \partial_z A_\zbar = 0 .
\]
In \autocite{CWY1, CWY2}, Costello, Witten, and Yamazaki used this gauge to analyze line defects in this four-dimensional gauge theory, as a systematic explanation for various quantum integrable systems. 
(We add further remarks about their work in Section~\ref{sec: CS revisited}.)

To treat this theory correctly, one needs to take into account the gauge invariance of this action functional,
and hence one can and should formulate it using the BV-BRST formalism. 
We do this carefully in the body of the text, and we will see how the gauge-fixing condition above translates to this setting. 

\begin{rmk}
More generally, this theory makes sense on a product $M \times X$ where $M$ is an oriented real 2-manifold and $X$ is a Riemann surface with a choice of holomorphic volume form $\omega$ (i.e., a global holomorphic section of the canonical line bundle $K_X$).
In that setting, the equation of motion picks out connections that are flat along $M$ and holomorphic along $X$.
In \autocite{CostelloYangian}, Costello explains how to extend to a larger class of spaces,
namely a complex surface $X$ with a reduced effective divisor $D$ and a trivialization of $K_X(2D)$,
but hopefully our description communicates the flavor of 4-dimensional Chern--Simons theory and where it might appear.
\end{rmk}

For this theory, our main result implies the following.

\begin{thm*}
On $\RR^2 \times \CC$, this 4-dimensional Chern--Simons theory admits a finite quantization at one loop.
\end{thm*}

In other words, by a judicious choice of gauge-fixing, 
we find that the one-loop Feynman diagrams have no UV divergences and the one-loop anomaly vanishes.
(In Section~\ref{sec: CS revisited} we discuss in more depth this result and how it relates to prior work.)

\def\ybar{\overline{y}}
\def\dybar{\d \ybar}

Moving up in dimension, there is a 5-dimensional Chern--Simons theory on $\RR \times \CC^2$ where both the $y$ and $z$ coordinates are complex.
In explicit terms, the gauge field is a $\fg$-valued 1-form of the following type:
\[
A = A_x \d x + A_{\ybar} \dybar + A_{\zbar} \dzbar
\]
where each coefficient $A_\bullet$ is a smooth $\fg$-valued function.
Note that any terms $A_y \d y$ and $A_z \d z$ are zero.
The action functional is
\[
S(A) = \int_{\RR \times \CC^2} \d y\, \d z \wedge \left( \frac{1}{2}\kappa\left(A, \d_{\RR} A + \dbar A \right) + \frac{1}{3}\kappa\left( A, [A,A]\right) \right),
\]
where
\[
\d_{\RR} A = \d x\frac{\partial A}{\partial x}  \quad\text{and}\quad
\dbar A  = \d \ybar\frac{\partial A}{\partial \ybar}  + \d \zbar \frac{\partial A}{\partial \zbar} .
\]
The equation of motion is then that
\[
\d_{\RR} A + \dbar A + \frac{1}{2} [A,A] = 0,
\]
which picks out a connection that is flat along $\RR$ and holomorphic along~$\CC^2$.
(As earlier, this type of theory makes sense on a larger class of 5-manifolds.)

For this theory, our main result implies the following.

\begin{thm*}
On $\RR \times \CC^2$, this 5-dimensional Chern--Simons theory admits a finite quantization at one loop.
\end{thm*}

In other words, by a judicious choice of gauge-fixing, 
we find that the one-loop Feynman diagrams have no UV divergences and the one-loop anomaly vanishes.
(Again, see Section~\ref{sec: CS revisited} for further discussion, particularly related to Costello's work on a noncommutative 5-dimensional Chern--Simons theory.)

\subsection*{Mixed BF theories}

Another important class of gauge theories are of BF-type.
Topological BF theory is a gauge theory on an $m$-dimensional 
smooth manifold $M$ (in this paper $M = \RR^m$) whose fundamental fields consist of a gauge field $A \in \Omega^1(M) \otimes \fg$ and a conjugate field $B \in \Omega^{m-2}(M) \otimes \fg^*$ ($B$ may be a twisted form if $M$ is unoriented) with the action functional
\[
S(A,B) = \int_M (B, F_A) = \int_M (B, \d A) + \frac{1}{2} \int_M (B, [A,A]).
\]
(To properly define the theory, one must use the BV-BRST formalism, but we postpone this discussion.
The term ``BF theory'' applies to a slightly more general geometric setup, but the description above is sufficient for our purposes.)
The equations of motion are
\[
F_A = \d A + \frac{1}{2}[A,A] = 0 \quad\text{and}\quad \d_A B = \d B + [A,B] = 0,
\]
and hence pick out a flat connection $\d_A$ and a covariantly-constant $(m-2)$-form $B$.
This theory shows up naturally throughout physics: in four dimensions, as a deformation of Yang-Mills theory (in a first order formulation); as a term in effective field theories for topological states of matter; and as a topological twist of supersymmetric gauge theories (as we discuss in the next subsection).

Holomorphic BF theory is a simple modification of topological BF theory.
Let $X$ be an $n$-dimensional complex manifold (in this paper, we consider only $X=\CC^n$), 
and let $A \in \Omega^{0,1}(X) \otimes \fg$ and $B \in \Omega^{n,n-2}(X) \otimes \fg^*$.
The action functional is
\[
S(A,B) = \int_X (B, \dbar A) + \frac{1}{2} \int_X (B, [A,A]).
\]
The equations of motion are
\[
\dbar A + \frac{1}{2}[A,A] = 0 \quad\text{and}\quad \dbar_A B = \dbar B + [A,B] = 0,
\]
and hence pick out a holomorphic connection $\dbar_A$ and a covariantly-holomorphic section~$B$.
This theory shows up naturally as a holomorphic twist of supersymmetric gauge theories.

Mixed BF theories exist on product manifolds $M \times X$ where $M$ is a smooth manifold of real dimension $m$ and $X$ is a complex manifold of complex dimension $n$.
The gauge field is a $\fg$-valued 1-form in which only $(0,\bu)$-forms appear along the $X$-direction,
so a gauge field $A$ is a sum $A_{10} + A_{01}$ where
\[
A_{10} \in \Omega^1(M) \widehat{\otimes}_\pi\, \Omega^0(X) \otimes \fg \quad\text{and}\quad A_{01} \in \Omega^0(M) \widehat{\otimes}_\pi\, \Omega^{0,1}(X) \otimes \fg.
\]
Similarly, the conjugate field $B$ only involves $(n,*)$-forms along the $X$-direction:
\[
B \in \Omega^{m-2}(M) \widehat{\otimes}_\pi\, \Omega^{n,0}(X) \otimes \fg^* \oplus \Omega^m(M) \widehat{\otimes}_\pi\, \Omega^{n,n-2}(X) \otimes \fg^*.
\]
The action functional is
\[
S(A,B) = \int_{M \times X} (B, \d_M A + \dbar_X A) + \frac{1}{2} \int_{M \times X} (B, [A,A]).
\]
The equations of motion are
\[
\d_M A + \dbar_X A + \frac{1}{2}[A,A] = 0 \quad\text{and}\quad \d_M B + \dbar_X B + [A,B] = 0,
\]
and hence pick out a connection that is flat along $M$ and holomorphic along $X$ as well as a covariant-constant section~$B$.

Our main result implies here the following.

\begin{thm*}
On $\RR^m \times \CC^n$ with $m>0$, mixed BF theory admits an exact and finite quantization at one loop.
\end{thm*}

In other words, by a judicious choice of gauge-fixing, 
we find that the one-loop Feynman diagrams have no UV divergences.
Moreover, no higher loop diagrams appear (for combinatorial reasons),
and the action functional satisfies the quantum master equation.

\subsection*{Twisted supersymmetric field theories}
\label{ex: susy twists}

Some of the examples above may appear as if they were engineered to fit into our context,
but they appear naturally as deformations of theories of broad interest.
In particular, every supersymmetric field theory produces THFTs through a process called {\em twisting}. 
Our theorems resemble some of the striking results about the perturbative behavior of supersymmetric theories,
and it would be interesting to understand how they are related.

By definition, a supersymmetric field theory on $\RR^d$ is a theory that is acted upon by a super Poincar\'{e} algebra. 
In this paper we restrict our attention Euclidean field theories and work in Riemannian signature,
hence for us the ``super Poincar\'{e} algebra'' is a super Lie algebra of the form
\[
\mathfrak{so}(n) \ltimes \ft
\]
where $\ft$ is the super Lie algebra of {\em supertranslations} whose even part $\ft^0 = \RR^d$ is the Lie algebra of ordinary translations and whose odd part $\ft^1$ is a sum of spin representations. 
While the Lie algebra $\RR^d$ of ordinary translations is abelian, the Lie algebra $\ft$ carries a nontrivial (super) Lie bracket,
and this Lie bracket is defined in terms of a $\mathfrak{so}(V)$-equivariant non-degenerate symmetric pairing
\[
\Gamma : {\rm Sym}^2(\ft^{1}) \to {\rm Sym}(\ft^0) \cong \RR^d 
\]
by the formula $[Q, Q'] = \Gamma(Q, Q')$. 

By a {\em supercharge}, one means an odd supertranslation $Q \in \ft$, and
a {\em twist} is a (nonzero) supercharge $Q$ such that $[Q,Q] = Q^2 = 0$. 
The classification of all twists is a completely algebraic question,
and we refer to~\autocite{ESsusy} for a complete classification of all twists in dimensions from $1$ to~$10$. 

If a Lagrangian field theory $\cT$ has $\ft$ as a symmetry, 
then a choice of twist $Q$ determines a deformation of the theory that we call a {\em twisted supersymmetric theory} $\cT^Q$.
If $S$ denotes the action functional of $\cT$, then the action functional $S^Q$ of $\cT^Q$ has the form
\[
S^Q(\varphi) = S(\varphi) + \int_{\RR^d} \varphi  \left(Q \cdot \varphi \right) + \cdots 
\]
where the deformation arises from how $Q$ acts on the theory 
(the $\cdots$ leaves room for terms non-linear in $Q$).
For extensive details on how the twisted theory is defined we refer to~\autocite{CostelloHolomorphic, ESW}. 

Elliott, Safronov, and the third author have given a complete characterization of the all twisted supersymmetric Yang--Mills theories in dimensions $2 \leq d \leq 10$ \autocite{ESW}. 
Twisted theories are THFT analogs of BF theory and Chern--Simons theory,
exactly the kinds of theories we have already discussed, so long as one includes purely topological and purely holomorphic theories. 
We will describe a few examples in detail in the BV formalism in Section~\ref{sec:twistedsusybv},
but we want to indicate now what our main theorem implies about some interesting examples.

For our purposes here, a crucial property of a twist $Q$ is the dimension $k_Q$ of its image ${\rm Im} \; \Gamma(Q, \cdot) \subset \RR^d$,
which we call the number of {\em invariant} directions of~$Q$.
By the non-degeneracy of the pairing $\Gamma$, it follows that $k_Q \geq \frac{d}{2}$.
If the number of invariant directions is maximal with $k_Q = d$, then the twisted theory $\cT^Q$ is purely topological. 
If the number of invariant directions is minimal with $k_Q = \frac{d}{2}$, then the twisted theory $\cT^Q$ is purely holomorphic. 

As a representative example of how twisting behaves, 
consider four-dimensional supersymmetric Yang--Mills theory with $\cN=2$ supersymmetry,
which means the odd part $\ft^1$ is eight-dimensional. 
We only consider here the pure gauge theory with Lie algebra~$\fg$. 
Then we have the bounds $2 \leq k_Q \leq 4$, and each value gives a different, well-known field theory:
\begin{itemize}
\item When $k_Q = 4$, the twisted theory is the topological field theory on $\RR^4$ known as {\it Donaldson--Witten theory} \autocite{WittenTQFT};
\item When $k_Q = 3$, the twisted theory is the THFT studied by Kapustin in \autocite{Kapustin};
\item When $k_Q = 2$, the twisted theory is holomorphic BF theory on $\CC^2$ with values in a graded Lie algebra $\fg[\ep]$, where $\ep$ is a parameter of cohomological degree~$1$.
\end{itemize}
We will describe these theories completely later in Section~\ref{sec:twistedsusybv},
but our main theorem implies the following.

\begin{thm*}
Donaldson--Witten theory, the Kapustin twist, and holomorphic BF theory for $\fg[\epsilon]$ each admit an exact and finite quantization at one loop.
\end{thm*}

In other words, by a judicious choice of gauge-fixing, 
we find that the one-loop Feynman diagrams have no UV divergences.
Moreover, no higher loop diagrams appear (for combinatorial reasons),
and the action functional satisfies the quantum master equation.
See Section~\ref{sec: susytwist revisited} for some further discussion.

In one dimension lower, we'd also like to point to work of Creutzig, Dimofte, Garner, and Greer who are studying minimal and topological twists of 3-dimensional $\cN=4$ gauge theory. 
The gauge-fixing condition we use here can be used to study such models at the quantum level. 

\subsection*{Applications and future directions} 

The results here overlap significantly with much other work (particularly on supersymmetric theories and their twists) but also open up a number of interesting directions of research.
We offer here a quick overview of our own such endeavors:
\begin{itemize}
\item Costello has utilized aspects of this gauge fixing condition to study twists of 11-dimensional supergravity in the $\Omega$-background \cite{CostelloM5}---the resulting theory is a noncommutative deformation of 5-dimensional Chern--Simons theory discussed above. 
Raghavendran, Saberi, and the third author use these techniques to study the quantization of twists of 11-dimensional supergravity (before turning on an $\Omega$-background).
\item The first and third authors use related techniques with Elliott on the Kapustin-Witten twists of 4-dimensional $\cN=4$ super Yang-Mills theory, to construct the factorization algebras of observables.
\item In the companion paper, we will study THFTs on more general manifolds; 
product manifolds $M \times X$ work, but there are more subtle examples, related to superconformal geometry.
\item The third author has developed techniques to extend the holomorphic gauge to complex manifolds (i.e., not just $\CC^d$), and we intend to develop the THFT analogs of such results, building on the global aspects of the companion paper.
\item The second author has developed a theory of renormalization of BV theories on manifolds with boundary so as to produce stratified factorization algebras~\cite{ERthesis}.
We intend to explore the applications of his methods to THFTs.
\item The three of us are working with Vicedo to study the THFT on $\RR \times \CC$ known as critical-level Chern-Simons theory, and its connections with the Gaudin model via chiral boundary conditions and line defects.
\end{itemize}

Many other possibilities beckon.

\subsection*{Acknowledgements}

This paper grew out of efforts to generalize and systematize the insights that led to \autocite{GWcs},
and so the origins of this work go back to conversations with Kevin Costello and Si Li.
At a technical level, we are deeply indebted to Li's work on holomorphic field theory.
In \autocite{LiZhou}, Li and Zhou have now developed a powerful new approach to 2-dimensional chiral CFT  that transcends Li's original work; 
we want to be a bit quicker absorbing this next bag of tricks!
In addition, our collaborations and interactions with Chris Elliott and Pavel Safronov have shaped how we understand THFTs, 
particularly how they relate to supersymmetric field theory.
E.R. would like to thank Benjamin Albert, from whom he first learned of these techniques of Costello and Li.
The National Science Foundation supported O.G. through DMS Grant No. 1812049.
E.R. is supported by the National Science Foundation Graduate Research Fellowship Program under Grant No. DGE 1752814. 
Work on this paper began in spring 2020, when all three authors benefited from the hospitality of the Mathematical Sciences Research Institute in Berkeley, California;
while O.G. and B.W. were in residence at MSRI, they received support from the NSF under Grant No. 1440140.
Any opinions, findings, and conclusions or recommendations expressed in this material are those of the authors and do not necessarily reflect the views of the National Science Foundation.

\subsection*{Notations and conventions}

Throughout this paper we will work on the manifold $\RR^m \times \CC^n$,
and we will work with linear coordinates $x_1,\ldots,x_m$ on $\RR^m$ and linear holomorphic coordinates $z_1,\ldots,z_n$ on $\CC^n$.
These coordinates equip the tangent bundles $T_{\RR^m} \to \RR^m$ and $T^{1,0}_{\CC^n}$ with canonical frames,
which we use throughout the text.
Similarly, we use $\d x_1, \ldots , \d x_m$ and $\d z_1 , \ldots, \d z_n$ as frames for the cotangent bundles.
We will use the notation $\d^m x$ to denote the volume form $\d x_1 \wedge \cdots \wedge \d x_m$ on $\RR^m$ and likewise the holomorphic volume $\d^n z = \d z_1 \wedge \cdots \wedge \d z_n$ on~$\CC^n$.
The operator $\d_{dR}$ denotes the de Rham differential acting on differential forms on $\RR^m$, while $\dbar$ denotes the anti-holomorphic Dolbeault operator acting on forms on~$\CC^n$. 

Recall that the completed projective tensor product $\potimes$ has the property that
\[
C^\infty(M) \potimes C^\infty(X) \cong C^\infty(M \times X),
\]
i.e., it corresponds to working with functions on the product manifold.
This feature extends, of course, to smooth sections of vector bundles, in the following way:
\[
\cinfty(M;E_1)\potimes \cinfty(X;E_2) \cong \cinfty(M\times X; E_1\boxtimes E_2),
\]
where $E_1\boxtimes E_2$ is the bundle on $M\times X$ whose fiber at a point $(m,x)$ is the tensor product $(E_1)_m\otimes (E_2)_x$.

We will denote by $A^\sharp$ the bundle of graded algebras on $\RR^m\times \CC^n$ given by the external tensor product $\Lambda^\sharp T^*_{\RR^m}\boxtimes \Lambda^\sharp (T_{\CC^n}^{0,1})^*$.
(We note that $A^\sharp$ is just graded and has no differential.)
The sheaf of sections of $A^\sharp$ is denoted $\cA^\sharp$, which is a sheaf of graded-commutative algebras via the wedge product.
When we equip it with the differential $\d_{dR} + \dbar$, we use the notation $\cA$ to denote this sheaf of dg commutative algebras.
(The bundle $A$ and the space $\cA$ have analogues when $\RR^m\times \CC^n$ is replaced with an arbitrary product $M\times X$, though we do not use this fact at all in the present work.)

We will adhere to a pattern of using normal-font Latin letters (e.g. $A$) for vector bundles on $\RR^m\times \CC^n$ and calligraphic font letters (e.g. $\cA$) for the corresponding sheaves of sections.

When $\cA$ is a sheaf on $\RR^m\times \CC^n$, we will often abuse notation and write $f\in \cA$ when we mean $f\in \cA(U)$ for some $U\subset \RR^m\times \CC^n$.

We also occasionally use subscripts to emphasize the space on which a sheaf resides. 
For example, $\Omega^\bullet_{\RR^m}$ refers to the sheaf of differential forms on $\RR^m$, and $\Omega^\bullet_{\RR^m}(U)$ refers to the value of this sheaf on an open subset $U\subset \RR^m$. 

By a {\em holomorphic} differential operator on $\RR^m \times \CC^n$, 
we mean a differential operator acting on $\cA$ but built from a holomorphic differential operator on $\CC^n$. 
Every holomorphic function on $\CC^n$ pulls back along the projection map to a function on $\RR^m \times \CC^n$,
so holomorphic functions form a subalgebra of smooth functions on $\RR^m \times \CC^n$.
Similarly, the vector fields of the form $f(z) \frac{\partial}{\partial z_i}$, where $f$ is holomorphic function on $\CC^n$,
form a sub-Lie algebra of the smooth vector fields on $\RR^m \times \CC^n$. 
If $D$ denotes the smooth differential operators on $\RR^m \times \CC^n$, 
the {\it algebra of holomorphic differential operators} denotes the smallest sub-algebra of $D$ containing the holomorphic functions and these vector fields.
These holomorphic differential operators have a canonical action on $\cA$, 
because they each commute with its differential.

Unless otherwise specified all functions on $\RR^m \times \CC^n$ are smooth.
By a slight abuse of notation we will often write the coordinate dependence of a smooth function as $G(x,z)$ referencing only the holomorphic coordinate in the $\CC^n$ direction.


\section{Definitions and examples} 
\label{sec:dfn}

In this section, we describe what we mean by a ``topological-holomorphic field theory'' on $\RR^m \times \CC^n$.
We have already gestured at concrete examples, but we need to formulate explicitly the full data necessary to define a field theory in the BV formalism.

\subsection{The definitions}

We begin with the case of free theories, and then indicate the allowed interaction terms for the case of interacting theories.
Costello and Gwilliam have developed an extensive formalism for studying perturbative quantum field theories in the BV formalism.
Our results will be proven within this formalism; hence, we will define a ``topological-holomorphic field theory'' as a field theory---in the sense of Costello and Gwilliam---which satisfies additional assumptions.


\subsubsection{Free THFTs}

\begin{dfn}
The {\em free topological-holomorphic field theory} associated to the data $(V, \<\cdot,\cdot\>_V, Q^{hol})$ where
\begin{itemize}
\item $V$ is a $\ZZ$-graded complex vector space equipped with a graded skew-symmetric non-degenerate pairing $\<\cdot,\cdot\>_V$ of degree $n+m-1$, and
\item a constant-coefficient holomorphic differential operator $Q^{hol}$ acting on $\cO(\CC^n) \otimes V$ of cohomological degree $+1$ satisfying $(Q^{hol})^2 = 0$,
\end{itemize} 
is the free BV theory on $\RR^m \times \CC^n$ given by the following data:
\begin{itemize}
\item[(1)] 
The fields are smooth sections of the graded vector bundle whose degree $p$ component is
\begin{equation}\label{eqn:Efree}
E^p = \bigoplus_{j+k+l = p} \Lambda^j T^*_{\RR^m} \otimes \Lambda^k T^{0,1 *}_{\CC^n} \otimes V^l = \bigoplus_{q+l = p} A^q \otimes V^l.
\end{equation}
The graded sheaf of fields $\cE$ is thus of the form
\[
\cE = \Omega^{\bullet}_{\RR^m} \, \widehat{\otimes}_\pi \, \Omega^{0,\bullet}_{\CC^n}  \otimes V = \cA \otimes V.
\]
\item[(2)] The differential on $\cE$ is given by
\begin{equation}\label{eqn:Qfree}
Q = \d_{dR} \otimes \id_{\Omega^{0,\bullet}_{\CC^n}} \otimes \id_V + \id_{\Omega^\bullet_{\RR^m}} \otimes \dbar \otimes \id_V + \id_{\Omega^\bullet_{\RR^m}}\otimes \id_{\Omega^{0,\bullet}_{\CC^n}} \otimes Q^{hol},
\end{equation}
or, more succinctly, $Q = \d_{dR} + \dbar + Q^{hol}$.
\item[(3)] The nondegenerate skew-symmetric pairing of degree~$-1$ 
\[
\ip_{loc} \colon E \otimes E \to \CC\, \d^m x \wedge \d^n z \wedge \d^n \zbar
\]
is defined by
\begin{equation}\label{eqn:ipfree}
\<\phi, \psi\>_{loc} = \<\phi, \psi\>_V  \wedge \d^n z  .
\end{equation}
\end{itemize}
The {\em action functional} of this free theory is
\[
S(\phi) = \int_{\RR^m \times \CC^n} \ip[\phi, Q \phi]_{loc}.
\]
\end{dfn}

The equation of motion of this theory is then $Q \phi = 0$, which expanded out is of the form
\[
\d_{dR} \phi + \dbar \phi + Q^{hol} \phi = 0 .
\]
When $Q^{hol} = 0$, this picks out sections of $E$ that are locally constant along $\RR^m$ and holomorphic along~$\CC^n$.
More accurately, the cohomology of the complex $(\cE,Q)$ of global sections of $E$ is $\cO(\CC^n) \otimes V$,
the $V$-valued entire functions, by the Poincar\'e and $\dbar$-Poincar\'e lemmas.

Another common choice of $Q^{hol}$ is of the form
\[
\frac{\partial}{\partial z_1} \d z_1  + \cdots + \frac{\partial}{\partial z_k} \d z_k
\]
for $1 \leq k \leq n$ and where the graded vector space $V$ is of the form $W \otimes\CC[ \d z_1, \ldots, \d z_k]$ for some graded vector space $W$. 
For $k = n$, this operator is precisely $\partial$,  and so $Q$ then amounts to the de Rham differential on $\RR^m \times \CC^n$.
In that case, the THFT can be viewed as living on $\RR^{m+2n} \times \CC^0$, i.e., be purely topological.
For $k < n$, the total operator $Q$ can be viewed as living on $\RR^{m + 2k} \times \CC^{n-k}$.
In other words, a theory does not have a preferred choice of THFT type:
a THFT on $\RR^m\times \CC^n$ is also a THFT on $\RR^{m-2}\times \CC^{n+1}$,
and we may freely trade pairs of topological directions for holomorphic directions. 
Sometimes the trade-off is beneficial, but we will see that our results are strongest for theories that possess at least one topological direction.

The following example shows that trading directions shows up even in fundamental examples, like Chern--Simons theory.

\begin{example}
The examples we mentioned in the introduction are all examples of interacting THFTs, which we will introduce in a moment,
but interesting free examples exist, as we now show.

There is a family of free 3-dimensional THFT defined on~$\RR \times \CC$
that includes both abelian Chern--Simons theory and a non-topological cousin. 
In other words, we will show how to view the usual Chern--Simons theory not as a purely topological theory but as living in a family of theories with THFT type $m = 1$ and $n =1$.

Set $V = \CC[1] \oplus \CC$, where the non-degenerate pairing is between the degree $-1$ and degree zero summand.
Define the differential operator $Q^{hol}$ acting on $\cO(\CC) \otimes V$ by 
\[
Q^{hol} \define k \frac{\partial}{\partial z} \colon \cO(\CC) \otimes \CC[1] \to \cO(\CC) \otimes \CC .
\]
where $k \in \CC$ is some constant.
Notice that $Q^{hol}$ is of degree~$1$ and is trivially square-zero. 

The complex of fields is 
\[
\cA[1]  \xto{k \frac{\partial}{\partial z}} \cA, 
\]
where $\cA = \Omega^\bu_\RR \widehat{\otimes}_\pi\, \Omega^{0,\bu}_\CC$.
Denote by $\alpha$ the sections of the component $\cA[1]$ and by $\beta$ the sections of $\cA$. 
The local pairing is then $\<\alpha, \beta\>_{loc} = (\alpha \wedge \beta) \wedge \d z$. 

If one absorbs the factor of $\d z$ into the definition of the field $\beta$, 
one can rewrite the complex of fields as
\[
\Omega^\bu_\RR \widehat{\otimes}_\pi\, \Omega^{0,\bu}_\CC [1] \xto{k \partial}  \Omega^\bu_\RR \widehat{\otimes}_\pi\, \Omega^{1,\bu}_\CC 
\]
where $\partial = \frac{\partial}{\partial z} \d z$ is the holomorphic de Rham operator. 
For any $k$, the fields then form the complex 
\[
\left(\Omega^\bu(\CC \times \RR) [1] , \dbar + \d_{\RR} + k \partial \right).
\]
Hence, when $k =1$, this complex is precisely the de Rham complex on $\RR^3$,
which is the BV description of ordinary abelian Chern--Simons theory on $\RR^3$.
When $k = 0$ and hence $Q^{hol} = 0$, it is an example of ``mixed'' BF theory that we will introduce in Section~\ref{sec:BF}. 
For $k \neq 0$, the theory is isomorphic to an ordinary abelian Chern--Simons theory, 
which can be seen by rescaling the $\Omega^{1,\bu}_\CC$ component of the complex above by the factor~$k^{-1}$.
\end{example}

\subsubsection{Interacting THFTs}

In the BV formalism, the action functional of an interacting theory must satisfy the classical master equation,
which leads to intricate relations between the homogeneous components of the interaction term (i.e., how the cubic term, quartic term, and so on fit together).
It can be convenient to rephrase these relations in terms of an $L_\infty$ algebra.
We will take that approach here, following the style of Section~4.4 of \autocite{CG2}.
For the reader who prefers action functionals, we offer that formulation as well.

First we briefly recall the $L_\infty$ algebra description of a BV theory.
It consists of the following data:
\begin{enumerate}
\item A $\ZZ$-graded vector bundle $L\to M$ on a manifold $Y$. 
We denote by $\cL$ the sheaf of sections of $L$.
\item For each $k\geq 1$, maps
\[
[\cdot,\ldots,\cdot]_k : \cL^{\otimes k} \to \cL[2-k]
\]
making $\cL$ into a sheaf of $L_\infty$ algebras on~$Y$.
These brackets are polydifferential operators.
\item A fiberwise non-degenerate vector bundle map
\[
\ip_{loc}: L\otimes L \to \textrm{Dens}_Y
\]
of $\ZZ$-degree --3.
The pairing is required to be invariant for the $L_\infty$ structure
so that the total data defines a sheaf of cyclic $L_\infty$ algebras on~$Y$.
\end{enumerate}

\begin{rmk}
\label{def: local lie}
Forgetting about the last piece of data, items (1)-(2) comprise the structure of a {\em local} $L_\infty$ algebra on $\cL$. 
`Local' here refers to the condition that all operations are polydifferential operators. 
\end{rmk}

The associated field theory arises by a shift of the above data,
namely, the fields are $\cE = \cL[1]$, which are sections of the bundle $L[1]$.
The action functional is
\[
S(\phi) = \sum_{n \geq 1} \frac{1}{(n+1)!} \ip[\phi, {[\phi, \ldots, \phi]_n}]_{loc},
\]
so that the $n$th summand is homogeneous of degree $n$.
In particular, the $n$-ary bracket determines the degree $n+1$ interaction term.
The equation of motion is
\begin{align*}
0 &= \sum_{n \geq 1} \frac{1}{n!} {[\phi, \ldots, \phi]_n} \\
&= [\phi]_1 + \frac{1}{2} [\phi,\phi]_2 + \frac{1}{3!}[\phi,\phi,\phi]_3 + \cdots,
\end{align*}
which can also be seen as the Maurer-Cartan equation of the $L_\infty$ algebra.

We now define THFTs in terms of further conditions on the brackets $[\cdot,\ldots,\cdot]_k$ and the pairing~$\ip_{loc}$.
We require a preliminary observation:
since we are working on the product manifold $\RR^m \times \CC^n$,
every holomorphic differential operator on $\CC^n$ determines a natural differential operator on $\RR^m \times \CC^n$.
Explicitly, if $\cD$ denotes the algebra of smooth differential operators acting on $\cA$,
let $\cD_{hol}$ denote the subalgebra generated by the Lie derivatives $L_{\del/\del z_k}$ and the holomorphic functions (i.e., those function on $\CC^n\times \RR^m$ which are pullbacks of holomorphic functions on~$\CC^n$).
We call these {\em holomorphic differential operators} on $\cA$.
Similarly, one can talk about holomorphic polydifferential operators on $\cA^{\otimes k}$.
More precisely, note that 
\[
\cA^{\hotimes_\pi k} \cong \cinfty_{\RR^{mk}\times \CC^{nk}}\otimes \CC[\d x^\alpha_i, \dzbar^\alpha_j]_{\alpha\in \{1,\ldots, k\}, i\in \{1,\ldots, m\}, j\in \{1,\ldots, n\}},
\]
i.e. the $k$-fold tensor product of $\cA$ with itself is an algebra of the same form but for different $m$ and $n$. 
A holomorphic polydifferential operator on $\cA^{\otimes k}$ is a polydifferential operator $\cA^{\hotimes_\pi k}\to \cA$ which arises as the composition of a holomorphic differential operator (in the above sense) on $\cA^{\hotimes_\pi k}$ with the map $\cA^{\hotimes_\pi k} \to \cA$ induced from the ``small diagonal''
\[
\RR^m\times \CC^n\ to (\RR^m\times \CC^n)^k.
\]
Finally, if $G$ is a vector bundle on $\RR^m\times \CC^n$, by a holomorphic polydifferential operator
\[
\Phi: \cG^{\hotimes_\pi k'}\hotimes_\pi \cA^{\hotimes_\pi k} \to \cA,
\] 
we mean a polydifferential operator such that, for fixed $g_1,\ldots, g_{k'}$, $\Phi(g_1,\ldots, g_{k'}, \cdot,\ldots, \cdot)$ is a holomorphic polydifferential operator.

\begin{dfn}
\label{dfn: thft}
A {\em strict topological-holomorphic field theory} on $\RR^m \times \CC^n$ is a classical BV theory on $\RR^m \times \CC^n$ given by the following data and conditions:
\begin{enumerate}
\item There is a free THFT determined by $(V, \<\cdot,\cdot\>_V, Q^{hol})$. Denote $L = E[-1]$ with $E$ as in Equation \eqref{eqn:Efree}. 
The pairing $\<\cdot,\cdot\>_{loc}$ is the one in Equation~\eqref{eqn:ipfree}. 
\item The $1$-ary map for $L$ is the operator $[\cdot]_1 = Q$ as in Equation~\eqref{eqn:Qfree}. 
\item For $k \geq 2$, the $k$-ary bracket on $\cL$ is a holomorphic polydifferential operator.
Together with $[\cdot]_1 = Q$, these brackets are required to endow $L = E[-1]$ with the structure of a sheaf of $L_\infty$ algebras. 
\item We require that the degree $(-3)$ pairing $\<\cdot,\cdot\>_{loc}$ on $L$ be invariant for this $L_\infty$-structure. 
\end{enumerate}
\end{dfn}

{\em Henceforth, we will use the abbreviation THFT to stand for ``strict topological-holomorphic field  theory.''}

In other words, a classical THFT is a classical BV theory whose underlying 

A number of remarks on this definition deserve to be made:
\begin{itemize}
\item In the case $m =0$, the definition of a THFT on $\RR^m \times \CC^n$ agrees with the definition of a holomorphic field theory on $\CC^n$ from~\autocite{LiVertex,BWhol}. 

\item As remarked earlier, a THFT on $\RR^m \times \CC^n$ can be viewed as a THFT on $\RR^{m-2k} \times \CC^{n+k}$ (But not necessarily the other way around).
\item This definition can (and should) be relaxed to allow polydifferential operators that are ``holomorphic up to homotopy.'' 
When one does this, one finds that each such non-strict theory is equivalent as a BV theory to a strict THFT, when the spacetime is $\RR^m \times \CC^n$.
(We will provide a careful discussion in a companion paper that focuses on global aspects of THFTs, where the spacetime can be more interesting.)
\item A careful examination of our proofs below show that our analytic bounds work if we allow brackets where the polydifferential operators can be built by wedging with elements of $\cA$ and by differentiating against the vector fields $\partial/\partial z_k$.
In essence the $\dbar$-Poincar\'e lemma indicates that such operators should be equivalent to holomorphic polydifferential operators.
\item So far, all examples of THFTs that we have encountered in ``nature'' (i.e., our research) are strict.
\end{itemize}

\subsection{Examples of THFTs}

We will revisit the theories discussed in the introduction, place them in the BV formalism, and then verify they are THFTs.

\subsubsection{$4$-dimensional Chern--Simons theory}
\label{sec:4dcs}

Consider $4$-dimensional Chern--Simons theory on $\RR^2 \times \CC$,
so that $m = 2$ and $n=1$.
The graded vector space $V$ is $\fg[1]$ and $Q^{hol}=0$. 
The non-degenerate, invariant, symmetric pairing $\kappa$ on $\fg$ defines a degree $(-1)$ density-valued pairing 
\[
\<\alpha \otimes X , \beta \otimes Y\>_{loc} = \d z \, (\alpha \wedge \beta)_{top}\, \kappa(X,Y) \in \CC \cdot \d z \, \d \zbar \, \d^2 x 
\]
for $\alpha, \beta \in \cA$ and $X,Y \in \fg$.
This completely describes the free THFT whose fields are~$\cA \otimes \fg[1] = \Omega^\bu_{\RR^2} \potimes \Omega^{0,\bu}_{\CC} \otimes \fg[1]$.  

The interacting THFT is defined by the following strict dg Lie algebra structure on $\cA \otimes \fg$ defined by
\[
[\alpha \otimes X, \beta \otimes Y] = (\alpha \wedge \beta) \otimes [X, Y] 
\]
for $\alpha, \beta \in \cA$ and $X,Y \in \fg$. 
Here the wedge product is of differential forms and the bracket is the Lie bracket on~$\fg$. 

The resulting BV action functional reads
\[
\frac12 \int_{\RR^2 \times \CC} \d z \, \kappa (A, \d A) + \frac{1}{6} \int_{\RR^2 \times \CC} \d z \, \kappa( A , [A, A]) = \int_{\RR^2 \times \CC} \d z \, {\rm CS}_\kappa (A) . 
\]
By contrast with the notation in the introduction, $A$ now stands for an arbitrary element of~$\cA \otimes \fg[1]$. 

\subsubsection{$5$-dimensional Chern--Simons theory}
\label{sec:5dcs}

Consider $5$-dimensional Chern--Simons theory on $\RR \times \CC^2$,
so that $m = 1$ and $n=2$.
The graded vector space $V$ is $\fg[1]$ and $Q^{hol}=0$. 
The non-degenerate, invariant, symmetric pairing $\kappa$ on $\fg$ defines a degree $(-1)$ density-valued pairing 
\[
\<\alpha \otimes X , \beta \otimes Y\>_{loc} = \d^2 z \, (\alpha \wedge \beta)_{top} \kappa(X,Y) \in \CC \cdot \d^2 z \, \d^2 \zbar \, \d x 
\]
for $\alpha, \beta \in \cA$ and $X,Y \in \fg$.
This completely describes the free THFT whose fields are~$\cA \otimes \fg[1] = \Omega^\bu_{\RR} \potimes \Omega^{0,\bu}_{\CC^2} \otimes \fg[1]$.  
The interacting THFT is defined by a strict dg Lie algebra structure on $\cA \otimes \fg$ defined just as in the preceding example. 
The resulting BV action functional reads
\[
\frac12 \int_{\RR \times \CC^2} \d^2 z \, \kappa( A , \d A) + \frac{1}{6} \int_{\RR \times \CC^2} \d^2 z \, \kappa( A , [A, A] ) =  \int_{\RR^2 \times \CC} \d^2 z \, {\rm CS}_\kappa (A). 
\]
By contrast with the notation in the introduction, $A$ now stands for an arbitrary element of~$\cA \otimes \fg[1]$.

\subsubsection{Topological-holomorphic BF theory}
\label{sec:BF}

Again, suppose $\fg$ is a Lie algebra, but do not fix an invariant pairing.
Then {\em BF theory} on $\RR^m \times \CC^n$ is an interacting THFT defined for any values~$m,n$.

In this case 
\[
V = \fg[1] \oplus \fg^* [n+m-2]\, \d^n z
\]
and $Q^{hol} = 0$. 
Define the density-valued local pairing by
\[
\<\alpha \otimes X + \alpha' \otimes X'\, \d^n z\, , \beta \otimes Y + \beta' \otimes Y' \, \d^n z\,\>_{loc} = \d^n z \, (\alpha \wedge \beta')_{top} \<X, Y'\>_\fg + \d^n z \, (\alpha' \wedge \beta')_{top}  \<X', Y\>_\fg .
\]
for $\alpha, \alpha',\beta,\beta' \in \cA$, $X,Y \in \fg$, and $X',Y' \in \fg^*$. 
Here $\<\cdot,\cdot\>_\fg$ stands for the evaluation pairing between $\fg$ and $\fg^*$. 
This completely describes the free THFT whose fields are 
\[
\cA \otimes (\fg[1] \oplus \fg^*[n+m-2]) .
\]

The interacting THFT is described by a strict Lie algebra structure on $\cA \otimes (\fg \oplus \fg^*[m+n-3])$ defined by
\begin{align*}
[\alpha \otimes X, \beta \otimes Y] & = (\alpha \wedge \beta) \otimes [X, Y] \in \cA \otimes \fg \\
[\alpha \otimes X, \beta' \otimes Y'] & = (\alpha \wedge \beta') \otimes [X,Y'] \in \cA \otimes \fg^* [n+m-3] \\
[\alpha' \otimes X', \beta' \otimes Y'] & = 0 
\end{align*}
for $\alpha, \alpha',\beta,\beta' \in \cA$, $X,Y \in \fg$, and $X',Y' \in \fg^*$. 
Here, $[\cdot, \cdot]$ denotes both the bracket in $\fg$ and the coadjoint action of $\fg$ on $\fg^*$. 

One can absorb the factor of $\d^n z$ in the definition of the pairing into the component $\fg^*$ of the fields to rewrite the fields as
\begin{equation}\label{eqn:bffields}
(A,B) \in \cA \otimes \fg[1] \oplus \cA \otimes \fg^*[m+n-2]   .
\end{equation}
Doing this, the action functional then reads
\[
S(A,B) = \int_{\RR^m \times \CC^n} \<B , F_A\>_\fg 
\]
where $F_A = (\dbar + \d_{dR}) A + \frac{1}{2} [A,A]$. 

\subsubsection{Twisted supersymmetric field theories}
\label{sec:twistedsusybv}

We now discuss a key source of interesting THFTs: supersymmetric field theories.
A much more extensive study of supersymmetric theories in the BV formalism is available in \autocite{ESsusy, ESW}, building on~\autocite{CostelloHolomorphic},
so we offer a mere sketch here as they pertain to THFTs.

Following the discussion from Section~\ref{ex: susy twists}, a supersymmetric theory for us will mean a classical BV theory on $\RR^d$ with an action of a super Lie algebra $\frak{so}(d) \ltimes \ft$.
That means, in particular, the cohomological vector field $\delta_S = \{S,-\}$ determined by the action functional commutes with the action of any element $v$ of the super Poincar\'e algebra.
Here we will focus on the subalgebra $\ft$ consisting of supertranslations (i.e., we ignore the $\frak{so}$-component).

A choice of twist $Q \in \ft$ determines both a deformation of the BV theory from $\cT$ to $\cT^Q$ (replacing $\delta_S$ with $\delta_S + Q$) and of the Lie algebra from $\ft$ to $\ft^Q = (\ft, Q)$.
As one might expect, these deformations are compatible, so that the dg Lie algebra $(\ft, Q)$ acts on the twisted theory.
Examining the twisted supertranslations, one finds that $Q$ encodes a choice of partial complex structure so that the Euclidean space can be seen as a product $\RR^m \times \CC^n$ and that the cohomology of the twisted translation algebra amounts to the Lie algebra generated by the vector fields $\frac{\partial}{\partial t_i}, \frac{\partial}{\partial z_j}, \frac{\partial}{\partial \zbar_j}$, for $i=1,\ldots m$ and $j = 1,\ldots, n$. 
Moreover, the action of $\ft^Q$ encodes that the translations $\frac{\partial}{\partial t_i}, \frac{\partial}{\partial \zbar_j}$, for $i=1,\ldots m$ and $j = 1,\ldots, n$ act in a homotopically trivial way. 
Such a homotopically trivial action ensures that the theory behaves like a topological theory along the $\RR^m$ directions and like a holomorphic theory along the $\CC^n$ directions.
The following result shows how nice the situation is. 

\begin{prp}
Let $\cT$ be a supersymmetric field theory on $\RR^d$ and let $Q$ be a nonzero supercharge satisfying $Q^2 = 0$. 
The twisted theory $\cT^Q$ is a translation-invariant THFT. 
\end{prp}

Here {\it translation-invariant} means that translations on $\RR^m \times \CC^n$ act as symmetries but also that translations along $\RR^m$ and anti-holomorphic translations along $\CC^n$ are homotopically trivial.

\begin{proof}
This result is completely algebraic and follows from Proposition 3.25 of \autocite{ESsusy}.
Let $k = {\rm dim} \left({\rm Im} \; \Gamma(Q, \cdot)\right)$ be the number of invariant directions. 
By the non-degeneracy of the pairing $\Gamma$, it follows that $k \geq \frac{d}{2}$.
Without loss of generality we can assume that $\frac{\partial}{\partial x_1} , \ldots, \frac{\partial}{\partial x_k}$ span ${\rm Im} \; \Gamma(Q, \cdot)$. 
It follows that we obtain a translation invariant THFT on $\RR^{2k-d} \times \CC^{d-k}$ with the remaining non-trivial holomorphic translations given by $\frac{\partial}{\partial z_i} = \frac{\partial}{\partial x_{k+i}}$ for $i = 1,\ldots, d-k$. 
\end{proof}
%

In \autocite{ESW} there is a catalogue of all twists of supersymmetric gauge theories in dimensions between 2 and 10,
so that there is a bounty of THFTs to study,
and they typically appear as topological-holomorphic BF theories and topological-holomorphic analogs of Chern--Simons theory.
We examine here one interesting example to exhibit our results.

Take the case of pure $\cN =2$ supersymmetric Yang--Mills theory on $\RR^4$.
As explained in Section~\ref{ex: susy twists}, there are three types of twists,
leading to a purely holomorphic theory, a mixed theory, and a purely topological theory.
(For more details on these characterizations of twists of four-dimensional supersymmetric gauge theory we refer to~\autocite[Section 10.2]{ESW}.)

The first case arises from a square-zero supercharge $Q_{0}$ satisfying the minimum bound for the number of invariant directions, which in this case is two. 
Associated to $Q_{0}$ there is a twist of the supersymmetric Yang--Mills theory that is equivalent to holomorphic BF theory on $\CC^2$ with values in a graded Lie algebra $\fg[\ep]$ where $\ep$ has degree $-1$.
This theory has fields
\begin{align*}
A + \ep A'& \in \Omega^{0,\bu}(\CC^2 , \fg[\ep])[1] \\
B + \ep B' & \in \Omega^{2,\bu}(\CC^2, \fg^* [\ep]) [1]
\end{align*}
and has action
\[
S_0 (A,A',B,B') = \int_{\CC^2} B \wedge \dbar A' + \int_{\CC^2} B' \wedge \dbar A + \frac12 \int_{\CC^2}  B' \wedge [A,A] + \int_{\CC^2} B \wedge [A, A']  ,
\]
where we implicitly use the canonical pairing between $\fg$ and $\fg^*$ to obtain a top form.

The second case arises from a square-zero supercharge $Q'$ that commutes with $Q_{\rm hol}$ and has the property that the sum $Q_{0} + Q'$ has {\em three} invariant directions. 
Indeed, we have a family of twisting supercharges $Q_{0} + u Q'$, where $u$ is a deformation parameter.
If we introduce holomorphic coordinates $(w,z) \in \CC^2$, 
these deformations of $Q_{0}$ amount to introducing a trivialization (up to homotopy) of the holomorphic derivative $\partial/\partial w$,
so that we work now on the space $\RR^2 \times \CC$ rather than $\CC^2$.
The associated twisted theories arise as a deformation of the holomorphic BF theory above,
where the family is described by the action functional
\[
S_u = S_{\rm 0} + u \int_{\CC^2} B \wedge \frac{\partial}{\partial w} A .
\]
For $u = 0$, this theory is holomorphic BF theory on $\CC^2$. 
For $u \ne 0$, this theory is isomorphic to the THFT version of BF theory on $\RR^2 \times \CC$ whose fields are
\begin{align*}
\Tilde{A} & \in \Omega^{\bu}(\RR^2) \otimes \Omega^{0,\bu} (\CC_z) \otimes \fg [1] \\
\Tilde{B} & \in \Omega^{\bu}(\RR^2) \otimes \Omega^{1,\bu}(\CC_z) \otimes \fg^* [1]
\end{align*}
and whose action is 
\[
\int_{\RR^2 \times \CC} \Tilde{B}\wedge  \dbar \Tilde{A} + \frac12 \int_{\RR^2 \times \CC} \Tilde{B} \wedge [\Tilde{A}, \Tilde{A}].
\] 
This theory was studied by Kapustin in~\autocite{Kapustin}.

The third case is given by a twist that has four invariant directions,
so that all translations are homotopically trivial.
It can be seen as a different deformation of the holomorphic twist, of the form $Q_0 + Q''$, and is typically called the {\em Donaldson--Witten twist}.
If $v$ denotes another deformation parameter, we can consider the full family of twisting supercharges
\[
Q_{u,v} = Q_0 + uQ' + v Q''.
\]
For generic $u,v$ the THFT resulting in the twist by the supercharge $Q_{u,v}$ will only be a $\ZZ/2$-graded BV theory. 
The family of action functionals corresponding to this family of twisting supercharges is of the form
\[
S_{u,v} = S_u + v \int_{\CC^2} B' \wedge A' = S_0 + u \int_{\CC^2} B \wedge \frac{\partial}{\partial w} A + v \int_{\CC^2} B' \wedge A'  .
\]
For $v \ne 0$ and for all $u$, this theory looks rather trivial: the only solution is that $A$ and $B$ vanish, 
because the complex underlying the free BV is contractible. 
This feature is a common feature of ``A-type twists,'' at least as formulated in this style, 
and is connected with the fact that we ought to work with the whole moduli of solutions (e.g., instantons or BPS states) rather than around a single solution.
Our techniques should be relevant in that setting but we defer a discussion to the future.

\subsection{Background fields: a generalization} 
\label{sec:back}

In this subsection, we discuss a generalization of the above-presented  formalism for THFTs. 
Namely, we introduce \emph{background fields} into the picture.  
The reader should feel free to skip this section until it becomes relevant to her;
this extension is useful for many applications but not needed to understand our key arguments.

Recall the notion of a local $L_\infty$ algebra from Remark~\ref{def: local lie},
which can encode background fields just as it can encode a BV field theory.
(One simply drops the requirement of a pairing.)

\begin{dfn}
\label{dfn: backgroundfields}
A {\em THFT with background fields} on $\RR^m \times \CC^n$  is the data:
\begin{itemize}
\item a THFT described by the local $L_\infty$ algebra $\cL$ with non-degenerate pairing $\<\cdot,\cdot\>_{loc}$,
\item a local $L_\infty$ algebra $\cG$, and
\item a local $L_\infty$ algebra structure on $\cG \oplus \cL$ that we denote by $\cG \ltimes \cL$
\end{itemize}
such that 
\begin{itemize}

\item[(1)] the exact sequence of sheaves of complexes
\[
0 \to \cL \to \cG \ltimes \cL \to \cG \to 0
\]
consists of (strict) maps of $L_\infty$ algebras,

\item[(2)] the $L_\infty$ structure on $\cG\ltimes \cL$ preserves the pairing $\<\cdot, \cdot\>_{loc}$ in the sense that for any collection of sections $\{\alpha_1,\ldots, \alpha_k\}$ of $\cG$ and $\{\beta_1,\ldots, \beta_{k+1}\}$ of $\cL$, then
\[
\<[\alpha_1, \ldots, \alpha_k, \beta_1, \ldots, \beta_\ell]_{k+\ell} , \beta_{\ell+1}\>_{loc}
\]
is graded totally anti-symmetric under permutation of the elements $\beta_i$, and

\item[(3)] for $k,k' > 0$, the brackets 
\[
[\cdot, \ldots, \cdot]_{k',k}:\cG^{k'}\times \cL^{k}\to \cL
\]
are holomorphic polydifferential operators with respect to the $\cL$ inputs, as described in the paragraph preceding Definition~\ref{dfn: thft}.
\end{itemize}
\end{dfn}
Note that as a consequence of condition (1), the only brackets $[\cdot]_k$ involving both $\cG$ and $\cL$ have the form
\[
\cG^{\times \ell} \times \cL^{\times \ell'} \to \cL
\]
where $\ell + \ell' = k$ with both $\ell$ and $\ell'$ positive. 

\begin{rmk}
This definition refines the definition of a field theory (not necessarily a THFT) with background fields from~\autocite{CG2}.
Definition \ref{dfn: backgroundfields} ensures that the underlying propagating field theory is a THFT. 
Moreover, Condition (3) of the definition ensures a further compatibility between---on the one hand---the action of the background fields $\cG$ on $\cL$ and---on the other---the topological-holomorphic structure of $\cL$.
Our theorems apply in the strongest sense only for THFTs with background fields as defined here.
We will use the expression ``$\cG$ is a symmetry of the THFT $\cL$'' when all conditions but Condition (3) of Definition \ref{dfn: backgroundfields} are met.
\end{rmk}
%
%
%
%

\subsection{Examples with background fields}

We now exhibit some useful examples.

\subsubsection{Charged matter}
\label{sec: chgdmatter}
In physics one often studies a field theory coupled to a background gauge field, 
such as electrons running around in an ambient, fixed electromagnetic field.
Here we describe a THFT example of this kind of situation, although we stick with bosonic ``matter.''
We work on $\RR^m \times \CC^n$. 

Choose a Lie algebra $\fg$.
We will construct a free THFT with background dg Lie algebra $\cG = \cA \otimes \fg$.
A degree 1 element of $\cG$ can be viewed as a gauge field, i.e., connection 1-form,
but it must determine a connection that is flat along $\RR^m$ and a $\dbar$-connection along~$\CC^n$.
(The dg Lie algebra also ensures we are working up to infinitesimal gauge equivalence.)

Next, choose a representation $P$ of $\fg$. 
For the field theory $\cL$, we take the following sheaf
\[
\left(\cA\otimes P[-1]\right) \oplus \left(\cA\otimes P^\ast[n+m-2]\right);
\]
we set $Q^{\rm{hol}}=0$, and $[\cdot, \ldots, \cdot]_k=0$ for $k\geq 2$.
For the pairing $\ip_{loc}$, we define
\[
\ip[\alpha_1\otimes p, \alpha_2 \otimes \rho]_{loc}=\rho(p) \, \alpha_1 \wedge \alpha_2\wedge \dz_1\wedge \cdots \wedge \dz_n .
\] 
(This theory $\cL$ is equivalent to BF theory, Section~\ref{sec:BF}, for the abelian graded Lie algebra $P[-1]$.)
The semi-direct product $\cG \ltimes \cL$ has a ``cross'' bracket $[\cdot, \cdot] = [\cdot, \cdot]_{1,1}$ defined by
\begin{align*}
[\beta \otimes X , \alpha \otimes p] & = (\beta \wedge \alpha) \otimes (X \cdot p)\\
[\beta \otimes X , \alpha \otimes \rho] & = (\beta \wedge \alpha) \otimes (X \cdot \rho)
\end{align*}
where $\alpha,\beta \in \cA$ and $X\in \fg$, $p \in P$, $\rho \in P^*$. 
Also, $X \cdot p$, $X \cdot \rho$ denote the action of $\fg$ on $P, P^*$ respectively.
The equations of motion for this theory encode sections that, with respect to the background gauge field, ``look flat'' along $\RR^m$ and ``look holomorphic'' along~$\CC^n$.
In this sense, the fields are ``charged'' because they are sensitive to an ambient gauge field.

There are many variants on this construction.
For instance, it is natural to replace $\cG$ by $\Omega^\bu \otimes \fg$, which encodes a global $\fg$-symmetry (this factors through the action of $\cA \otimes \fg$ we just described via the natural projection~$\Omega^\bu \to \cA^\bu$ of commutative differential graded algebras).

\subsubsection{Holomorphic vector fields and mixed BF theory}

Consider the following local $L_\infty$ algebra, which encodes some of the ``spacetime symmetries'' relevant to THFTs. 

\begin{example} 
\label{eg:holvf}
Let $W = {\rm span}_\CC \left\{\frac{\partial}{\partial z_1}, \ldots, \frac{\partial}{\partial z_n}\right\}$ and consider the cochain complex 
\[
\cG = \cA \otimes W .
\]
As usual, the differential is $\dbar + \d_{dR}$.
There is a Lie bracket defined by
\[
\left[\alpha(z,\zbar, x) \frac{\partial}{\partial z_i}, \beta(z,\zbar, x) \frac{\partial}{\partial z_j}\right] = \alpha \left(\frac{\partial}{\partial z_i} \beta\right) \frac{\partial}{\partial z_j} - (-1)^{|\alpha| |\beta|} \beta \left(\frac{\partial}{\partial z_j} \alpha\right)\frac{\partial}{\partial z_i}.
\]
It is straightforward to check that the differential $\delbar+\d_{dR}$ is a derivation for this Lie bracket.
This defines the structure of a topological-holomorphic local dg Lie algebra on $\cG$ that we refer to as the local dg Lie algebra of {\em holomorphic vector fields on~$\RR^m \times \CC^n$.}

A more invariant description of this dg Lie algebra goes as follows.
Let $T_{\CC^n}$ denote the holomorphic tangent bundle on $\CC^n$. 
Then $\Omega^{0,\bu}_{\CC^n} (T_{\CC^n})$ has the structure of a dg Lie algebra, which uses the $\dbar$-connection on $T_{\CC^n}$ and the commutator of vector fields for the Lie bracket. 
Since $\Omega^\bu_{\RR^m}$ is a commutative dg algebra, the tensor product 
\[
\cG \cong \Omega^\bu_{\RR^m} \potimes \Omega^{0,\bu}_{\CC^n}(T_{\CC^n})
\]
is also a dg Lie algebra. 
\end{example}

Now consider holomorphic BF theory on $\RR^m \times \CC^n$ with fields $\cL$ labeled $(A,B)$ using the presentation in Equation~\eqref{eqn:bffields}.
It has a natural action by $\cG$, the holomorphic vector fields just defined.
The total space of fields of the THFT with background fields is $(X; A,B) \in \cG \oplus \cL$ where $\cG \oplus \cL$ is 
\[
\Omega^{\bu}_{\RR^m} \potimes \Omega^{0,\bu}(\CC^n, T_{\CC^n}) \oplus \left(\cA \otimes \fg \oplus \cA \otimes \fg^*[m+n-3] \right)  .
\]
The Lie algebra structures on $\cL$ and $\cG$ are defined as above. 
The only new piece of data is in defining the Lie algebra structure on $\cG \ltimes \cL$.
There is a single new bracket of the form
\[
[\cdot,\cdot]_{1,1} \colon \cG \times \cL \to \cL
\]
defined by
\[
[\alpha\otimes X , A] = \alpha\wedge L_X A , \qquad [\alpha\otimes X ,B] = \alpha\wedge L_X B
\]
where $\alpha\in \cA$, $X$ is a smooth $(1,0)$ vector field, and $L_X(\cdot)$ denotes the Lie derivative. 
It is immediate to see that this bracket preserves the local pairing on~$\cL$.

\subsubsection{Background fields in two-dimensional topological BF theory}
\label{sec: bfexamples}

In this section, we consider a variety of examples of symmetries of topological BF theory to concretely illustrate the nature of THFTs with background fields.

As we have remarked above, a topological theory in two dimensions can also be understood as a holomorphic theory in one complex dimension, and BF theory is no exception.
(See \autocite{bfascft, LMYonBF} for a work that takes advantage of this perspective.)
We will discuss three symmetries of BF theory below.
The first two symmetries do {\em not} result in THFTs with background fields.
The third example results in a THFT with background fields only if one understands BF theory as a holomorphic theory (but not as a topological theory).

\begin{example}
\label{ex: holvecbf}
Let $\cL$ be the local Lie algebra underlying BF theory on $\RR^2$, which is just ordinary topological BF theory, so
\[
\cL = \Omega^{\bu}_{\RR^2} \otimes \fg \oplus \Omega^\bu_{\RR^2} \otimes \fg^*[-1] .
\]
Let $\cG$ be the Lie algebra of smooth vector fields on $\RR^2$, so 
\[
\cG = \mathrm{Vect}_{\RR^2}  .
\]
The total space of fields, including background fields, is spanned by elements of the form $(X ; A,B) \in \cG \oplus \cL$.
The sheaf $\cG$ is canonically a sheaf of differential graded Lie algebras, as is $\cL$.
It remains to define a differential graded Lie algebra structure on $\cG\oplus \cL$.
Such a structure is determined by the following map~$[\cdot, \cdot] = [\cdot, \cdot]_{1,1} \colon \cG \times \cL \to \cL$ defined by
\begin{equation}
[X,A] = L_X A
\qquad
[X,B] = -L_X B.
\end{equation}
Altogether, these data determine a symmetry of topological BF theory.
Note, however, that since the action of a vector field on the space of fields involves (non-holomorphic) derivatives, this symmetry does not satisfy condition (3) of Definition \ref{dfn: backgroundfields}, hence does not furnish a THFT with background fields.
\end{example}

The following non-example of a background field plays a role in defining BF theory on arbitrary two-manifolds,
although we will not explore that aspect here. 
It involves a general construction of wide use.

\begin{dfn}\label{dfn:trivial}
If $\fh$ is a (ordinary) Lie algebra, let $\HH \fh$ denote the dg Lie algebra whose underlying cochain complex is concentrated in degrees $(-1)$ and zero and given by
\[
\fh[1] \xto{{\rm id}} \fh.
\]
The Lie bracket on $\HH \fh$ extends the one on $\fh$ by the adjoint action $\fh \times \fh[1] \to \fh[1]$. 
If $(M,\d)$ is a dg $\fh$-module,
a {\em homotopical trivialization} of the $\fh$ action is a lift of $\fh \to {\rm End}(M)$ to a map of dg Lie algebras $\HH \fh \to {\rm End}(M)$. 
\end{dfn}

The following example extends the previous one.
In so doing, it illustrates another way a symmetry may fail to furnish a THFT with background fields.

\begin{example}
\label{ex: nonexample}
As in the preceding example, we take $\cL$ to be the differential graded Lie algebra corresponding to topological BF theory on~$\RR^2$.
Consider the Lie algebra ${\rm Vect}_{\RR^2}$ of smooth vector fields and define 
\[
\cG = \HH {\rm Vect}_{\RR^2}.
\]
In the previous example, we saw that $\cL$ is a module for ${\rm Vect}_{\RR^n}$, and in fact ${\rm Vect}_{\RR^n}$ acts by symmetries on $\cL$.
We observe that $\cL$ is a trivializable module by extending the dg Lie algebra structure on $\cG \oplus \cL$ from the previous example by the formulas
\begin{equation}
\label{eq: hmtpyforrotation}
[ X,A] = \iota_X A
\qquad\text{and}\qquad
[X,B] = -\iota_X B,
\end{equation}
where $X$ is a degree $-1$ element in $\HH {\rm Vect}_{\RR^2}$.
By Cartan's formula $[\d, \iota_X] = L_X$, we have described the structure of a local differential graded Lie algebra on $\cG\oplus \cL$.
One further verifies that Conditions (1) and (2) of Definition~\ref{dfn: backgroundfields} are satisfied.

Hence $\cG $ acts by symmetries on topological BF theory but does not satisfy the conditions to provide a THFT with background fields.
Indeed, as we have already seen, the action of $\mathrm{Vect}_{\RR^2}$ does not satisfy Condition (3) of Definition \ref{dfn: backgroundfields} because it involves taking derivatives of fields in directions in which the theory is supposed to be topological. 
But Condition (3) of Definition \ref{dfn: backgroundfields} is further violated because of the interior multiplication in Equation~\ref{eq: hmtpyforrotation}:
Condition (3) forbids brackets that do not act as the identity on the $\CC[\d x_j, \dzbar_i]$ parts of~$\cA$.
\end{example}

Finally, we present an example of a symmetry of BF theory that yields a THFT with background fields only if one considers BF theory to be a holomorphic theory on~$\CC$.
This symmetry will exhibit some interesting subtleties about our results.

\begin{example}\label{ex: semiexample}
Consider the Lie algebra $\textrm{Vect}^{1,0}_\CC$ of {\em smooth} vector fields of type $(1,0)$ on $\CC$.
That is, consider vector fields of the form $f(z,\zbar) \frac{\partial}{\partial z}$ for $f(z,\zbar)$ some smooth function.
Since it is a Lie subalgebra of the algebra of all smooth vector fields on $\RR^2$,
$\HH \textrm{Vect}^{1,0}_\CC$ acts by symmetries on topological BF theory by Example~\ref{ex: nonexample}, which we just discussed.
If one understands topological BF theory to be a THFT on $\RR^2$ (i.e. a purely topological theory), then this symmetry does {\em not} furnish a THFT with background fields, for the same reasons as in Example \ref{ex: nonexample}.
However, if we view topological BF theory on $\RR^2$ as a holomorphic theory on $\CC$,
then the formulas above {\em do} have the structure of a THFT with background fields. Consider the local Lie algebra $\HH \textrm{Vect}^{1,0}_\CC \oplus \cL$
with the same Lie brackets but with $Q^{hol} = \partial = \d z \frac{\partial}{\partial z}$ now.

When one considers BF theory as a holomorphic theory, the $\d z$ in the space of fields is absorbed into the ``flavor'' of the theory, i.e., into the space $V$ appearing in Definition~\ref{dfn: thft}; further, since we are considering only vector fields of type (1,0), the action of $\cG$ on $\cL$ only involves holomorphic derivatives of fields, which \emph{is} allowed in condition (3) of Definition \ref{dfn: backgroundfields}.
It follows that the symmetry $\HH \mathrm{Vect}^{1,0}_\CC$ of 2-dimensional BF theory does furnish a THFT with background fields when BF theory is considered as a holomorphic theory on~$\CC$.
\end{example}

\section{The topological-holomorphic gauge}
\label{sec: gf}

In this section we turn to the key technical tool involved in the proof of each of our main results, the \emph{topological-holomorphic gauge}. 
Recall that, in the formalism of Costello, one of the main inputs for the quantization of a gauge theory is a gauge-fixing operator (whose definition is recalled below).
The topological-holomorphic gauge is a special gauge that one may define for any THFT on $\RR^m \times \CC^n$. 
In this section, we study this gauge in detail.

\begin{dfn}
Let $\cE$ be a classical BV theory whose underlying free action is of the form $\int \<\phi, Q \phi\>_{loc}$. 
A \emph{gauge-fixing operator} $Q^{\rm GF}$ for $\cE$ is a degree $-1$ differential operator on $\sE$ satisfying
\begin{itemize}
\item The (anti)-commutator $[Q,Q^{GF}]$ is a generalized Laplacian (in the sense of \autocite{bgv}).
\item The equation $(Q^{GF})^2=0$ holds.
\item The operator $Q^{GF}$ is graded symmetric for the pairing $\ip$.
\end{itemize}
\end{dfn}

Consider a THFT on $\RR^m \times \CC^n$. 
Recall, the space of fields is of the form $\cA \otimes V$, with $V$ some graded vector space. 
To construct a gauge fixing operator we will take advantage of an obvious product metric on $\RR^m \times \CC^n$. 

\begin{dfn}
\label{def: mixedgf}
The {\em topological-holomorphic gauge-fixing operator} for a THFT is the operator
\[
Q^{GF}:=\d_{dR}^\star \otimes \id_{\Omega^{0,\bullet}}\otimes \id_V + 2 \, \id_{\Omega^\bullet} \otimes \delbar^\star\otimes \id_V,
\]
where $\d_{dR}^\star$ and $\dbar^\star$ are the (formal) adjoints of $\d_{dR}$ and $\dbar$ for the flat Euclidean metric on $\RR^n$ and flat K\"{a}hler metric on~$\CC^n$, respectively. 
\end{dfn}

\begin{lmm}
The operator defined in Definition \ref{def: mixedgf} is a gauge-fixing operator for any THFT.
\end{lmm}

\begin{proof}
This is just a statement about free THFTs. 
The underlying free theory for any THFT is described by a differential on the complex $\cE = \cA \otimes V$ of the form
\[
Q = \d_{dR} + \dbar + Q^{hol}
\]
with $Q^{hol}$ a holomorphic differential operator acting on $\cinfty_{\RR^m\times \CC^n}\otimes V$. 
By definition
\[
[\d_{dR}, \dbar^\star] = [\dbar, \d_{dR}^\star] = [Q^{hol} , \d_{dR}^\star + \dbar^\star] = 0 .
\]
The last equality follows from the fact that $Q^{hol}$ is a constant-coefficient holomorphic operator. 
From this characterization it follows that the only pairs of operators appearing in $Q$ and $Q^{GF}$ which don't commute are $\delbar, \delbar^\star$ and $\d_{dR},\d_{dR}^\star$.
Hence,
\[
[Q,Q^{GF}] = [\d_{dR},\d_{dR}^\star] + 2[\delbar, \delbar^\star],
\]
which is readily recognized to be the standard Laplacian on Euclidean space. 
The other two properties of a gauge-fixing operator are easily verified.
\end{proof}

\begin{rmk}
We again emphasize the possibility that a given theory might have different descriptions as a mixed theory. 
Hence, the operator $Q^{GF}$ in Definition \ref{def: mixedgf} is not uniquely associated to a THFT; 
rather, it is defined by a decomposition of the theory into holomorphic and topological directions. 
In particular, we may always view a THFT on $\RR^{m+2} \times \CC^n $ as a THFT on $\RR^m \times \CC^{n+1}$. 
(Here, the underlying spacetimes are canonically isomorphic; 
we are simply emphasizing in which directions the theory is allowed to have a holomorphic nature.) 
Sometimes this maneuver represents a trade-off. 
Using the gauge obtained from the knowledge that a theory is topological in some directions improves the convergence of Feynman diagrams. 
On the other hand, the structure of the theory may be simpler when it is expressed as having more holomorphic directions. 
The first and third named authors leveraged this trade-off in \autocite{GWcs}. 
Namely, they showed when one considers Chern--Simons as a THFT on $\RR\times\CC$ (as opposed to one on $\RR^3 \times \CC^0$), one finds that the theory is one-loop exact. 
On the other hand, the Feynman diagrams do not \emph{a priori} converge, and one must do a bit of work to show that they do.
One of the consequences of this work is to show that all THFTs admit finite one-loop quantizations.
\end{rmk}

Given a gauge-fixing operator, we follow \autocite{CosBook} and define what we mean by a quantization of a field theory. 
It is out of the scope of this paper to review all aspects of this definition here, but we will simply provide a brief gloss of the main ingredients in Costello's definitions, both to orient the reader and to fix notation. 

One starts with the data of a classical BV theory whose action is of the form
\begin{equation}\label{eqn:classical action}
S(\phi) = \int \<\phi, Q \phi\>_{loc} + I(\phi)
\end{equation}
where $I(\phi) = \sum_{k \geq 2} \frac{1}{k!} \int \<\phi, [\phi \cdots \phi]_k]\>_{loc}$. 
Additionally, one fixes a gauge fixing operator $Q^{GF}$. 

Because $[Q,Q^{GF}]$ is a generalized Laplacian, it possesses a heat kernel $K_L$ which can be used to define the scale $L$ {\em BV Laplacian} $\Delta_L$ (not to be confused with the generalized Laplacian) and the scale $L$ {\em BV bracket} $\{\cdot, \cdot\}_L$ on the space $\cO(\cE)$ of functions on the space of fields. 
One then forms the equation 
\[
QI + \frac{1}{2}\{I, I\}_L +\hbar \Delta_L I=0,
\]
for elements of $\cO(\cE)[\![\hbar]\!]$, which is the \emph{scale $L$ quantum master equation (QME)}. 

One defines the RG flow operator $W(P_{\epsilon<L}, \cdot)$ on $\cO(\cE)$, 
which takes solutions of the scale $L$ QME to solutions of the scale $\epsilon$ QME. 

Costello defines a {\em quantum field theory} as a collection $\{I[L]\}_{L>0}$ of elements of $\cO(\cE)[\![\hbar]\!]$ satisfying 
\begin{enumerate}
\item the equation $W(P_{\epsilon<L}, I[L]) = I[\epsilon]$ (homotopical RG flow equation),
\item an asymptotic locality condition as $L\to 0$, and
\item the quantum master equation (QME)
\begin{equation}\label{eqn:anomaly}
QI[L]+\frac{1}{2}\{I[L],I[L]\}_L + \hbar\Delta_L I[L]=0
\end{equation}
for some (and hence, all) $L>0$.
\end{enumerate}
Let us focus initially on the first two conditions. 

\begin{dfn}
A collection $\{I[L]\}$  satisfying only conditions (1) and (2) above is called an {\em effective family of interactions}. 
\end{dfn}

A na\"{i}ve way that one can obtain an effective family is by using the RG flow operator $W(P_{\epsilon<L}, \cdot)$. 
When $0 < \epsilon < L$, the RG flow operator is completely well-defined. 
One generally runs into trouble when $\epsilon \to 0$, and this operator may not exist.

\begin{dfn}
\label{dfn: fte}
A classical field theory equipped with a gauge fixing operator $Q^{GF}$ is called {\em finite} if the limit
\[
\lim_{\epsilon \to 0} W (P_{\epsilon<L} , I) 
\]
exists. 
\end{dfn}

If a classical field theory is finite then one obtains an effective family by the rule $I[L] = \lim_{\epsilon \to 0}W(P_{\epsilon<L}, I)$. 
General field theories rarely satisfy this finiteness condition, however, as this $\epsilon \to 0$ limit is often ill-defined. 
In general, Costello describes a (non-canonical) procedure for obtaining an effective family $\{I[L]\}$ from a classical field theory. 
The essence of this procedure is to study divergences appearing in na\"{i}ve Feynman diagrams, and to systematically ``subtract off'' these divergences using ``counterterms".

Using the topological-holomorphic gauge, we will show that for a THFT on $\RR^m \times \CC^n$ the na\"{i}ve effective family is {\em finite} to first order in $\hbar$ (see Theorem \ref{thm: nodivs}).
Hence, there is no need to introduce counterterms and an effective family is obtained by running the na\"ive RG flow from scale zero to scale~$L$. 

By general maneuvers with homotopical RG flow, one sees that if an effective family satisfies condition (3) at a single scale $L$ then it holds at {\em all} scales.
If condition (3) does not hold we say that the effective family is {\em anomalous} with scale $L$ {\em obstruction} given by the left-hand side of Equation~\eqref{eqn:anomaly}.

We return to condition (3) and the QME in Section \ref{sec:anomaly}. 
When $m\geq 1$, we show that this effective family associated to a THFT satisfies the QME up to order $\hbar$ in Theorem \ref{thm: noanomaly}. 
In other words, to first-order in $\hbar$ the obstruction to solving the QME vanishes. 
(The case $m=0$ is studied in~\autocite{BWhol} where the same conclusion is false---holomorphic theories are plagued by one-loop anomalies.)  

As an immediate corollary of Theorems \ref{thm: nodivs} and \ref{thm: noanomaly}, we will find that mixed BF theory on $\RR^m \times \CC^n$, $m \geq 1$, possesses an exact one-loop quantization without the need to introduce any counter terms. 

In the remainder of this section, we set up the requisite background needed to execute the proofs of the aforementioned results.

\subsection{Integral kernels}

We proceed to construct the integral kernels for the heat operator and the (mollified) propagator,
and we point out some important properties they possess.
In Section~\ref{sec: nodivs}, we then examine one-loop diagrams and verify that no counterterms are needed.

As a first step in simplification, notice that $Q$ and $Q^{GF}$ are of the form $D\otimes \id_{V}$ for $D$ a differential operator on~$\cA$.
Hence the propagator factors as 
\[
P_{\epsilon<L} = P^{an}_{\epsilon<L} \otimes P^{V},
\]
where $P^{an}_{\epsilon<L}\in \cinfty((\RR^m\times \CC^n)^2)$ and $P^V\in V\otimes V$.
We will thus focus here only on the analytic component.

In the remainder of this section we set
\[
Y = \RR^m \times \CC^n
\]
and let $\Ypunc = Y \setminus \{0\}$,
the punctured vector space.
We use $(x_1,\ldots, x_m,z_1,\ldots, z_n)$ as the standard linear-holomorphic coordinates on~$Y$.

Let $\Conf_2(Y)$ denote $Y \times Y \setminus {\rm Diag}$, 
where Diag denotes the diagonal copy $Y \hookrightarrow Y \times Y$, 
as it is the space of ordered configurations of two points in~$Y$.
We equip it with coordinates 
\[
\left(x^1_1,\ldots, x^1_m, z^1_1,\ldots, z^1_n; x^2_1,\ldots, x^2_m, z^2_1,\ldots , z^2_n \right),
\]
where the semi-colon separates the coordinates on the first and second copies of $Y$ in $Y\times Y$. We will often abbreviate this as
\[
\left( x^1,z^1; x^2, z^2\right),
\]
letting $x^1$ stand for the whole collection $x^1_1,\cdots, x^1_m$, and so on.

There is a natural map $p\colon Y\times Y \to Y$ sending a pair of coordinates to their difference, 
as $Y$ is a vector space.
We will often be interested in the restriction of $p$ to the configuration space,
namely $p\colon \Conf_2(Y) \to \Ypunc$. 
We will find that pullback of forms along $p$ plays a crucial role in analyzing our integral kernels.

\begin{rmk}
On flat space the explicit formulas for $\delbar^\star$ and $\d_{dR}^\star$ are
\[
\delbar^\star = 2 \sum_{i=1}^n\iota_{\del/\del_{\zbar_i}}\frac{\partial}{\partial z_i} \quad\text{and}\quad \d_{dR}^\star = \sum_{j=1}^m\iota_{\partial/\partial x_j}\frac{\partial }{\partial x_j},
\]
where we have made use of the standard coordinates on $\RR^m\times \CC^n$.
\end{rmk} 


We have seen that the Laplacian arising from this gauge fixing operator,
\begin{equation}
\label{def of lap}
H = [Q,Q^{GF}] = 2 [\delbar,\delbar^\star] + [\d_{dR},\d_{dR}^\star],
\end{equation}
is the standard Laplacian on flat space.

We wish now to construct and analyze the heat kernels and mollified propagators arising from $Q$ and $Q^{GF}$.
It is convenient to introduce the following notation.
These integral kernels are distributional differential forms ({\em aka} currents) on $Y \times Y$ that are smooth on $\Conf_2(Y) = Y \times Y \setminus {\rm Diag}$.

The heat kernel for this Laplacian {\em acting on functions} is
\[
k^{an}_T(x^1,z^1;x^2,z^2) = \frac{1}{(4 \pi T)^{\frac{2n+m}{2}}} e^{-|x^1-x^2|^2/4T} e^{-|z^1-z^2|^2/4T}
\]
at {\em length scale} $T > 0$.
The heat kernel for this Laplacian {\em on the sections} $\cA$ is
\[
K^{an}_T = k^{an}_T \prod_{j=1}^m(\d x^1_j - \d x^2_j)  \wedge \prod_{i=1}^n (\d \overline{z}^1_i - \d \overline{z}^2_i).
\]
Note that we have simply multiplied the function $k^{an}_T$ by a differential form,
which is the pullback of a form on the diagonal $Y$ along the difference map $Y \times Y \to Y$ along the diagonal.
Hence, let's write 
\begin{align*}
\mu_Y = \prod_{j=1}^m\d x_j \wedge \prod_{i=1}^n\d \overline{z}_i,
\end{align*}
to introduce some compact notation.
Then $K^{an}_T = p^*( k^{an}_T  \mu_Y)$, which is more concise.

We now obtain the mollified propagator
\begin{equation}
\label{def of P}
P^{an}_{\epsilon<L} (x^1,z^1; x^2, z^2) = \int_{T=\epsilon}^L \d T \, Q^{GF} K_T^{an}(x^1,z^1; x^2, z^2) .
\end{equation}
Using the definition of $Q^{GF}$, we obtain
\begin{equation}
\label{eq: PfromE}
P^{an}_{\epsilon<L} (x^1,z^1; x^2, z^2)  = -\int_{T=\epsilon}^L (p^* E_T)(x^1,z^1; x^2, z^2)  ,
\end{equation} 
where $E_T$ is the $(m+n-1)$-form on $Y$ defined by
\begin{align}
\label{def of diag P}
E_T&(x,z) 
\define Q^{GF} (k^{an}_T\mu_Y) \\
&= 
-\frac{e^{-(|z|^2 + |x|^2)/4T}}{4(4 \pi)^{\frac{2n+m}{2}}T^{\frac{2n+m+2}{2}}} 
\left(\sum_{j=1}^{n} 
\left( \overline{z}_j\prod \d x_i \prod_{k\neq j}\d \overline{z}_i\right)
+\sum_{l=1}^m \left(x_l \prod_{q\neq l} \d x_q \prod \d \overline{z}_p \right)
\right) 
\end{align}

The two terms correspond to removing the line element $\d \overline{z}_i$ or $\d x_i$ from $\mu_Y$ followed by taking the associated derivative of the heat kernel.


In other words, we remove the line element $\d \overline{z}_i$ or $\d x_i$ from $\mu_Y$ and 
take the associated derivative of the heat kernel.
It will be convenient to keep track of those summands as
\begin{equation}
\label{E decomp}
E_T = E^{\d}_T + E^\delbar_T
\end{equation}
where the term $E^{\d}_T$ arises from applying $\d^\star_{dR}$ and $E^\delbar_T$ arises from applying~$\delbar^\star$.

It will also be convenient to introduce the simpler Gaussian expression
\begin{equation}
\label{def of G}
G_T(x,z) \define \frac{e^{-(|z|^2 + |x|^2)/4T}}{(4\pi T)^{\frac{2n+m}{2}}}.
\end{equation}
Note that we obtain $E_T$ from $G_T$ by applying a differential operator:
\begin{equation}
\label{rel of G and E}
E_T(x,z) = 
\left( 2\sum_{j=1}^n \prod_{i} \d x_i \prod_{k\neq j}\dzbar_k\frac{\partial}{\partial z_j} 
+ \sum_{l=1}^m\prod_{q\neq l} \d x_q \prod_p \dzbar_p \frac{\partial}{\partial x_l} \right)G_T(x,z).
\end{equation}
We denote this constant-coefficient differential operator by~$\lambda$.

\subsection{Some computational results}

There are a number of computational results that we will use repeatedly, 
and they are quite clever and helpful.
These are modest generalizations of work by Si Li, starting with \autocite{LiFeynman, LiVertex}, so we refer to them as {\em Li's tricks}. 
(In their recent paper \autocite{LiZhou}, Li and Zhou have introduced a whole new bag of fascinating tricks,
and it would be wonderful to see how they might appear in the setting of THFTs as well.)

We need to know how the vector field $\partial/\partial z_i$ acts on the propagator and the heat kernel.
Direct computation shows that
\begin{equation}
L_{\partial/\partial z_i} P^{an}_{\epsilon<L} (x^1,z^1; x^2, z^2) = \int_{T=\epsilon}^L  p^*\left(\frac{\zbar_i E_T}{4T} \right)\d T
\end{equation}
and hence that every further holomorphic derivative multiplies the integrand by a factor~$\zbar/4T$.
We note, given a multi-index $I$,
\begin{equation}
\label{d_z of P}
\frac{\partial^I}{\partial z^I} P^{an}_{\epsilon<L} (x^1,z^1; x^2,z^2) = (-1)^{|I|}\int_{T=\epsilon}^L p^*\left( \frac{ \zbar_I E_T(x,z)}{(4T)^{|I|}}\right) \d T;
\end{equation}
we will hereon use derivatives (without the Lie derivative notation).


The following notations will be used when we study the graph integral of a wheel $\gamma$ with $k$ vertices.

Let us denote by $(r^1,\ldots, r^{k})$ the standard coordinates on product manifold $Y^k$ where $r^\alpha = (x^\alpha_1,\ldots x^\alpha_m, z^\alpha_1,\ldots, z^\alpha_n)$.

\subsubsection{Center of mass coordinates}\label{sec:com}

We introduce the ``center of mass'' coordinates $(q^1,\ldots,q^{k})$ on $Y^k$ as follows:
\begin{equation}
\label{eq: coords1}
q^\alpha = r^{\alpha+1}-r^\alpha
\end{equation} 
for $1 \leq \alpha <k$,
and 
\begin{equation}
\label{eq: coords2}
q^{k} = r^{k}.
\end{equation} 
For each $\alpha$ write $q^\alpha = (y^\alpha_1,\ldots, y^\alpha_m, w^\alpha_1,\ldots, w^\alpha_n)$.
So, for instance, $y^1_1 = x^2_1 - x^1_1$. 
Generally, the $y^\alpha$ are coordinates on $\RR^m$ and the $w^\alpha$ are holomorphic coordinates on $\CC^n$, defined in terms of the $x^\alpha$ and $z^\alpha$, respectively, in the same way that the $q^\alpha$ are defined in terms of the~$r^\alpha$.

Consider the product 
\begin{equation}
\label{prod of Gs}
G_{\vec{T}}^{(k)}(q^1,\ldots,q^{k-1}) \define \prod_{\alpha=1}^{k-1} G_{T_\alpha}(q^\alpha) \cdot G_{T_{k}}\left(\sum_{\alpha=1}^{k-1} q^\alpha\right),
\end{equation}
where $\vec{T} = (T_1,\ldots,T_k) \in [\epsilon,L]^{k}$. 
We define $E^{(k)}_{\vec T}(q^1,\ldots, q^{k-1})$ analogously.

{\em Important observation}\/: Note that while $\vec{T} = (T_1,\ldots, T_k)$ is a $k$-tuple, 
the expression $G_{\vec{T}}^{(k)}(q^1,\ldots,q^{k-1})$ only depends on the center of mass variables $q_1,\ldots, q_{k-1}$. It is independent of~$q_k$.

More precisely, for $\alpha=1,\ldots, k-1$, we define 
\[
p_\alpha: (\RR^m\times \CC^n)^k\to \RR^m \times \CC^n
\]
by projection onto the coordinate $q^\alpha$; for $\alpha =k$, we define
\[
\begin{array}{cccc}
p_{k}:& (\RR^m\times \CC^n)^k &\to &\RR^m \times \CC^n\\
& (q^1,\ldots, q^{k}) & \mapsto & \sum_{\alpha=1}^{k-1} q^\alpha,
\end{array}
\]
and
\begin{equation}
\label{eq: prodofEs}
E^{(k)}_{\vec T}(q^1,\ldots, q^{k}) = \prod_{\alpha=1}^{k} p_\alpha^*  E_{T_\alpha}
\end{equation}
Direct computation shows the following.

\begin{lmm}
\label{lmm: zeta}
The differential operator
\[
\zeta^i = \frac{1}{\sum_{\alpha=1}^k T_\alpha} \sum_{\alpha = 1}^{k-1} T_\alpha \frac{\partial}{\partial w_i^\alpha}
\]
acts on the Gaussian form $G_{\vec{T}}^{(k)}$ by
\[
\zeta^i (G^{(k)}_{\vec{T}}) = -\frac{\sum_{\alpha = 1}^{k-1} \wbar_i^\alpha}{4T_k} G^{(k)}_{\vec{T}}.
\]
\end{lmm}

\begin{proof}
Compute that
\begin{equation}
\label{d/dz_jE}
\frac{\partial}{\partial w_i^\alpha} G^{(k)}_{\vec{T}} = -\frac{1}{4}\left(\frac{\wbar_i^\alpha}{T_\alpha} + \frac{\sum_{\alpha = 1}^{k-1} \wbar_i^\alpha}{T_k}\right) G^{(k)}_{\vec{T}}
\end{equation}
by the product rule applied to~\eqref{prod of Gs}. 
Now plug into the definition of~$\zeta^i$.
\end{proof}

\begin{crl}
\label{clever diff op}
For $1 \leq i < k$, 
\[
\left( \frac{\partial}{\partial w_i^\alpha} - \zeta^i\right) G^{(k)}_{\vec{T}} = -\frac{\wbar_i^\alpha}{4T_\alpha}G^{(k)}_{\vec{T}}.
\]
\end{crl}

For $\partial/\partial y_j$, we have  analogous results.

\begin{lmm}
\label{tau}
The differential operator
\[
\tau^j = \frac{1}{\sum_{\alpha=1}^k T_\alpha} \sum_{\alpha = 1}^{k-1} T_\alpha \frac{\partial}{\partial y_j^\alpha}
\]
acts on the Gaussian form $G_{\vec{T}}^{(k)}$ by
\[
\tau^j(G^{(k)}_{\vec{T}}) = -\frac{\sum_{\alpha = 1}^{k-1} y_j^\alpha}{2T_k} G^{(k)}_{\vec{T}}.
\]
\end{lmm}

\begin{crl}
\label{clever diff op 2}
For $1 \leq  \alpha < k$, 
\[
\left( \frac{\partial}{\partial y_j^\alpha} - \tau^j\right) G^{(k)}_{\vec{T}} = -\frac{y_j^\alpha}{2T_\alpha}G^{(k)}_{\vec{T}}.
\]
\end{crl}
As a matter of notation, we will write 
\[
\frac{\partial}{\partial w^\alpha}-\zeta
\]
for the collection of operators
\[
\frac{\partial}{\partial w^\alpha_i}-\zeta^i
\]
as $i$ ranges over $1,\ldots, n$. 
We apply a similar notation for the $y^\alpha$ and $\tau^i$.

We now wish to relate $G^{(k)}_{\vec{T}}$ to the corresponding object $E^{(k)}_{\vec{T}}$ built out of the $E_{T_\alpha}$.
By Equation~\eqref{rel of G and E}, we see that
\begin{equation}
\label{rel of big G and E}
E_{\vec{T}}^{(k)}(q^1,\ldots,q^{k}) = \prod_{\alpha=1}^{k}\lambda_\alpha G^{(k)}_{\vec{T}},
\end{equation}
where, for $1\leq \alpha \leq k-1$, 
\[
\lambda_\alpha = \sum_{j=1}^m c_{j \alpha}\left(  \frac{\partial}{\partial y_j^\alpha}-\tau^j\right) + \sum_{i=1}^n d_{i\alpha} \left(\frac{\partial}{\partial w_i^\alpha}-\zeta^i\right),
\]
and
\[
\lambda_k = \sum_{j=1}^m c_{j \alpha}\left(\tau^j\right) + \sum_{i=1}^n d_{i\alpha} \left(\zeta^i\right).
\]
Here, the $c_{j\alpha}$'s are forms of Dolbeault degree $n$ and de Rham degree $m-1$ and the $d_{i\alpha}$'s are of Dolbeault degree $n-1$ and de Rham degree~$m$.

%

\subsection{Recollections about Gaussian integrals}
\label{sec on big gaussian}

Finally, consider the Gaussian measure:
\begin{equation}
\label{big gaussian}
\mu(q^1,\ldots,q^{k-1}) =  
\exp\left( - \sum_{\alpha=1}^{k-1} \frac{|q^\alpha|^2}{4T_\alpha} - \frac{1}{4T_{k}}\left|\sum_{\alpha=1}^{k-1} q^\alpha\right|^2\right)
\prod_{\alpha=1}^{k-1}\d q^\alpha=\prod_\alpha\left( T_\alpha^{m/2+n} \d q^\alpha\right) G^{(k)}_{\vec T}.
\end{equation}
Here, each $dq^\alpha$ is the Lebesgue measure on the corresponding copy of~$\RR^m\times \CC^n$.

Integrals against this measure, particularly moments, can be quickly computed in terms of the quadratic form appearing in the exponential.
The matrix associated to the quadratic form is
\begin{equation}
M_{\vec{T}} =
\begin{pmatrix}
a_1+b & b & b & \cdots & b \\
b & a_2+b & b & \cdots & b \\
b & b & a_2+b & \cdots & b \\
\vdots & \vdots & \vdots & & \vdots \\
b & b & b & \cdots & a_{k-1}+b
\end{pmatrix}
\label{eq: Mdefn}
\end{equation}
where $a_\alpha = 1/T_\alpha$ for $1 \leq \alpha \leq k-1$ and $b = 1/T_{k}$.
As the matrix is the sum of an invertible diagonal matrix $A = \mathrm{diag} (a_1,\ldots, a_{k-1})$ and a rank one matrix $B$ (the all $1/4T_k$ matrix),
the Sherman-Morrison theorem \autocite{SM} gives an explicit formula for the inverse matrix:
\[
M_{\vec{T}}^{-1} = \begin{pmatrix}
c_1 + d_{11} & d_{12} & d_{13} & \cdots & d_{1, k-1} \\
d_{21} & c_2+ d_{22} & d_{23} & \cdots & d_{2, k-1} \\
d_{31} & d_{32} & c_3 + d_{33}& \cdots & d_{3,k-1} \\
\vdots & \vdots & \vdots & & \vdots \\
d_{k-1,1}& d_{k-1, 2} & d_{k-1, 3}& \cdots & c_{k-1}+ d_{k-1,k-1}
\end{pmatrix}
\]
where 
\[
c_\alpha = T_\alpha  \;\;\; , \;\;\; d_{\alpha\beta} = d_{\beta\alpha} = - \frac{T_{\alpha} T_\beta}{T_1 + \cdots + T_k}
\]
for $1 \leq \alpha,\beta\leq k-1$. 

The inverse determinant is
\begin{equation}
\label{matrix det}
\det\left( M_{\vec{T}}^{-1}\right) = \frac{T_1 T_2 \cdots T_k}{T_1 + T_2 + \cdots + T_k}
\end{equation}
by the matrix determinant lemma.

These formulas makes computing moments straightforward,
as moments can be expressed in terms of entries of the inverse matrix.
For instance, it will be important later to understand the $T$-dependence of moments.
A standard computation shows that 
\[
\int_{Y^{k-1}} y_{i}^{\alpha}y_{j}^{\beta} \mu \propto \left( \frac{T_1 T_2 \cdots T_k}{T_1 + T_2 + \cdots + T_k} \right)^{m/2+n} \frac{T_\alpha T_\beta}{T_1 + \cdots + T_k}
\]
if $\alpha \neq \beta$ and
\[
\int_{Y^{k-1}} y_{i}^{\alpha}y_{j}^{\beta} \mu \propto \left( \frac{T_1 T_2 \cdots T_k}{T_1 + T_2 + \cdots + T_k} \right)^{m/2+n} \left(T_\alpha - \frac{T_\alpha^2}{T_1 + \cdots + T_k} \right)
\]
if $\alpha = \beta$.
(The power $m/2+n$ since each entry in our matrix $M$ corresponds to whole copy of~$\RR^m \times \CC^n$.)
By the expressions above we mean that the right hand side is proportional to the left hand side by a coefficient which is independent of $\vec{T}$. 

For convenience later, we record some key features of these moments,
which follow from standard computations with Gaussians.
Notationally, given $\nu = (n^1_1, \ldots, n^1_m, n^2_1, \ldots, n^{k-1}_m)$ in $\NN^{m(k-1)}$,
let $y^\nu$ denote the monomial~$\prod_{\alpha = 1}^{k-1} \prod_{i=1}^m (y^\alpha_i)^{n^\alpha_i}$.
Note that the moment vanishes unless the total degree $|\nu| = n^1_1 + \cdots + n^{k-1}_m$ is even.

\begin{lmm}
\label{lmm std gauss}
Given $\nu \in \NN^{m(k-1)}$ of even total degree, 
the $T$-dependence of the moment $\int_{Y^k} y^\nu \mu$ is a sum of terms of the form
\[
\frac{T^\lambda}{(T_1 + \cdots + T_k)^{|\lambda|/2}} \frac{1}{(T_1 + \cdots + T_k)^{m/2+n}}
\]
where the exponent $\lambda \in \NN^k$ satisfies
\begin{itemize}
\item if $y^\alpha_i$ divides $y^\nu$, then $T_\alpha$ divides $T^\lambda$ and
\item there is equality $|\lambda| = |\nu|$ of total degrees.
\end{itemize}
\end{lmm}

\section{There are no divergences}
\label{sec: nodivs}

Our main technical result is the following.

\begin{thm}
\label{thm: nodivs}
The topological-holomorphic gauge fixing operator of any THFT on $\RR^m \times \CC^n$ (with or without background fields) produces an effective family which is finite modulo~$\hbar^2$ (cf. Definition~\ref{dfn: fte}) .
\end{thm}

The proof of this theorem involves some explicit calculus, which we perform in this section. 
It remains to check that the effective family (modulo $\hbar^2$) obtained using Theorem \ref{thm: nodivs} satisfies the quantum master equation.
That is the object of Section~\ref{sec:anomaly} and Theorem~\ref{thm: noanomaly}.

\subsection{The ingredients and some simplifications}
\label{weights and reduction}

Let us make some preliminary remarks on the combinatorial structure of the Feynman diagrams relevant to Theorem~\ref{thm: nodivs}.
The sum in $W(P_{\epsilon,L}, I)$ is over connected graphs constructed from the following ingredients:
\begin{enumerate}
\item All vertices have two types of incident half-edges: propagating half-edges and background half-edges.
\item All vertices have valence at least three.
\item Internal edges may be formed only from propagating half-edges.
\end{enumerate}
To each vertex, one assigns a term from the interaction $I$; 
to each internal edge, one assigns a propagator $P_{\epsilon<L}$.
For any classical field theory, the terms appearing at order $\hbar^0$ in $W(P_{\epsilon<L},I)$ are finite;
these correspond to trees.
As we are considering $W(P_{\epsilon<L}, I)$ modulo $\hbar^2$,
we only need to study the weights assigned to diagrams with precisely one loop.
Any connected diagram with one loop consists of a wheel---i.e. a one-loop diagram that remains connected when any of its edges is severed---attached to any number of trees. 
(In language more common to physicists, a wheel is a one-loop, one-particle irreducible diagram.)
As we have discussed, trees do not contribute to divergences in $W(P_{\epsilon<L}, I)$;
it therefore suffices to prove Theorem \ref{thm: nodivs} by studying only wheel diagrams.

Let us introduce some notation.
The definition of the weight of a wheel involves placing the propagator at each internal edge and the interaction $I$ at each vertex. 
The weights are evaluated by placing compactly supported fields 
\[
\varphi \in \cG_c[1]\oplus \cL_c[1] = \cG_c[-1]\oplus \cA_c \otimes V
\]
at each of the external edges.

We will make use of two simplifications:
\begin{enumerate}
\item First, the only $\epsilon$ dependence appears in the analytic part of the propagator $P_{\epsilon<L}^{an}$, 
so we may assume $V=\CC$ and $\cG=\cinfty_{\RR^m\times \CC^n}$.
In other words, we assume that all external edges are labeled by compactly supported forms in 
$\cinfty_{\RR^m\times \CC^n,c}\oplus \cA$
\item Second, each vertex, determined by a term in $I$, corresponds to an integral of the form
\begin{equation}
\int_{\RR^m \times \CC^n} D'_1 X\, \cdots D'_{k'} X\, \left( D_1(\varphi)\wedge \cdots \wedge D_k(\varphi) \right) \d^n z,
\end{equation}
where $X\in \cG$, $\phi\in \cA$, and the $D_i$ and $D'_i$ are holomorphic differential operator (i.e., only involve $\frac{\partial}{\partial z_i}$-derivatives). 
Some of the differential operators will hit the compactly supported forms placed on the external edges of the graph.
The remaining operators will hit the internal edges labeled by the propagators.
Since a holomorphic differential operator preserves the space of compactly supported functions, 
we may absorb the derivatives acting on external inputs into the inputs themselves.
\item For any holomorphic differential operator of the form 
\[
f(z_i) \frac{\del}{\del z_{i_1}}\cdots\frac{\del}{\del z_{i_l}},
\]
we may also absorb the factor $f(z_i)$ into the external field inputs.
\end{enumerate}
Thus, for the $\epsilon \to 0$ behavior it suffices to look at weights of wheels with arbitrary compactly supported functions as inputs where each of the internal edges are labeled by some {\it translation-invariant} holomorphic differential operator
\begin{equation}
D = \sum_{p_1,\ldots p_n} \frac{\partial^{p_1}}{\partial z_{1}^{p_1}}\cdots \frac{\partial^{p_n}}{\partial z_{n}^{p_n}}
\end{equation}
applied to the propagator $P_{\epsilon<L}^{an}$.
The following definition captures the essential analytic information relevant for this section. 

\begin{dfn}\label{dfn: analytic weight}
Let $\epsilon , L > 0$. 
In addition, fix the following data.
\begin{enumerate}
\item An integer $k \geq 1$ encoding the number of vertices along the wheel.
\item For each $\alpha = 1, \ldots, k$ a sequence of natural numbers
\begin{equation}
\vec{p}^\alpha = (p_1^\alpha, \ldots, p_n^{\alpha}) .
\end{equation}
We denote by $(\vec{p}) = (p^{\alpha}_i)$ the corresponding $n \times k$ matrix of integers. 
\end{enumerate}
The {\em analytic weight} associated to the pair $(k, (\vec{p}))$ is the smooth distributional form
\begin{equation}
W_{\epsilon < L}^{k, (\vec p)} \colon \cinfty_c((\RR^m\times\CC^n)^k)[\d x^\alpha_i, \dzbar^\beta_j] \to \CC,
\end{equation}
that sends a smooth compactly supported form 
\[
\Phi \in\cA((\RR^m\times\CC^n)^k) = C_c^\infty(\RR^{mk} \times \CC^{nk})[\d x^\alpha_j,\dzbar^\beta_i]
\]
to the value $W_{\epsilon < L}^{k, (\vec p)} (\Phi)$ given by the integral
\begin{equation}
\label{weight1}
\displaystyle \int_{(x^1,z^1,\ldots,x^k, z^k) \in (\RR^m \times \CC^n)^k} \left(\prod_{\alpha=1}^k \d^n z^\alpha\right)
\Phi(x^1,z^1, \ldots,z^k,x^k) \prod_{\alpha = 1}^k \left(\frac{\partial}{\partial z^\alpha}\right)^{\vec{p}^\alpha} P_{\epsilon < L}^{an}(p^{\alpha}, p^{\alpha+1}) .
\end{equation}
In the above expression, we use the convention that~$p^{k+1} = p^1$.
\end{dfn}

As in the previous section, set $Y = \RR^m \times \CC^n$. 
The coordinates on $Y^k$ are given by $\{r_i^\alpha=(x_j^\alpha, z_i^\alpha)\}$ where $\alpha = 1,\ldots,k$, $i = 1, \ldots, n$, and $j= 1,\ldots, m$. 
For each $\alpha$, $\{z_1^\alpha, \ldots, z_d^\alpha\}$ is a coordinate system for the space $\CC^n$ sitting at the vertex labeled by $\alpha$ and similarly for the $x_j^\alpha$.
Denote by $\pi_\alpha \colon Y^k \to Y$ the map 
\[
\pi_\alpha (r_1,\ldots, r_k) = r_\alpha - r_{\alpha+1} .
\]
with the cyclic convention~$x_{k+1} = x_1$. 
We also used
\begin{equation}
\left(\frac{\partial}{\partial z^\alpha}\right)^{\vec{p}^\alpha} = \frac{\partial^{p^\alpha_1}}{(\partial z^\alpha_1)^{p_1^\alpha}} \cdots  \frac{\partial^{p^\alpha_n}}{(\partial z^\alpha_n)^{p^\alpha_n}}
\end{equation}
as a shorthand notation.

We will refer to the collection of data $(k, (\vec{n}))$ in the definition as {\em wheel data}
because the weight $W_{\epsilon < L}^{k, (\vec p)}$ is the analytic part of the full weight $W_{\Gamma}(P^V_{\epsilon<L}, I)$ where $\Gamma$ is a wheel with $k$ vertices. 

To prove Theorem~\ref{thm: nodivs}, it thus suffices to show that the $\epsilon \to 0$ limit of the analytic weight $W_{\epsilon < L}^{k, (\vec{p})}(\Phi)$ exists for any choice of wheel data $(k, (\vec{p}))$.
To do this, there are two steps. 
First, we show the weights vanish when $k \leq m+n$ for purely algebraic reasons. 
Second, when $k > m + n$, we show the weights have finite $\epsilon\to 0$ limit by producing explicit bounds;
these areguments are the most technical aspect of the section.  
From these two results, Theorem~\ref{thm: nodivs} follows immediately. 

A finer-grained view of the analytic weights will be useful in both arguments, 
so we now introduce a further decomposition of~$W_{\epsilon < L}^{k, (\vec{p})}(\Phi)$.
Recall that each analytic propagator has the form 
\[
P^{an}_{\epsilon<L}(p^\alpha, p^{\alpha+1}) = -\int_\epsilon^L \d T_\alpha\, E_{T_\alpha}
\]
where
\[
E_{T_\alpha} = E_{T_\alpha}^\dbar + E_{T_{\alpha}}^\d
\]
as shown in~\eqref{E decomp}.
Thus, in taking the product over $\alpha$ of the propagators in~\eqref{weight1},
we can make binomial expansion
\[
W_{\epsilon < L}^{k, (\vec{p})}(\Phi) = \sum_{S \subset \underline{k}} W_{\epsilon < L}^{k, (\vec{p}), S}(\Phi) 
\]
where $S$ runs over subsets of $\underline{k} = \{1, 2,\ldots, k\}$ and where $S$ labels the edges $\alpha \in S$ on which the term $E_{T_\alpha}^\d$ is used (the term $E_{T_{\alpha}}^\dbar$ is placed on each edge in the complement~$S^c$).
Explicitly, we have that the $S$-weight is given by
\begin{equation}
\label{weight S}
W_{\epsilon < L}^{k, (\vec{p}), S}(\Phi)  
= \displaystyle \int_{\vec T \in [\epsilon, L]^{k}} \d\vec T \int_{Y^k} \left(\prod_{\alpha=1}^k \d^n z^\alpha\right)\Phi(x,z) \left(\prod_{\alpha \in S} D_{\alpha} \pi_\alpha^*E_{T_\alpha}^{\d} \prod_{\beta \in S^c} D_\beta \pi_\beta^*E_{T_\beta}^\dbar \right)
\end{equation}
where the $D_\alpha$ are constant-coefficient holomorphic differential operators.

In the remainder of this section we refer to the ``center of mass'' coordinates $(y^\alpha, w^\alpha)$ introduced in Section \ref{sec:com}.

Expanding out the weight, one finds that the integrand of the 
takes the form
\[
\Phi(x,z)\,  P^S\left (\frac{y^1}{T_1},\ldots,\frac{y^{k-1}}{T_{k-1}},\frac{\sum_{\alpha=1}^{k-1}y^\alpha}{T_k},\frac{\wbar^1}{T_1},\ldots,\frac{\wbar^{k-1}}{T_{k-1}},\frac{\sum_{\alpha=1}^{k-1}\wbar^\alpha}{T_k}\right)
G^{(k)}_{\vec T}
\]
because the term arising from the propagator becomes a Gaussian $G^{(k)}_{\vec T}$ multiplied by a polynomial $P^S$ in the listed variables.
Each factor of the form $\frac{\wbar^\alpha_{i_\alpha}}{T_\alpha}$ appearing in $P^S$ comes from the application to the propagators of the operator $\frac{\partial}{\partial z_i}$ in $\dbar^\star$ or of the holomorphic derivatives appearing in the interactions.
Similarly, each factor of the form $\frac{y^\alpha_{i_\alpha}}{T_\alpha}$ in $P$ comes from the application of the operator $\frac{\partial}{\partial x^\alpha_i}$ in $\d_{dR}^\star$ applied to the propagator. 
Note that $\dbar^\star$ commutes with any translation-invariant holomorphic differential operator, 
so it does not matter in which order we do this.

It will be convenient to decompose $P$ further by treating the real and complex directions distinctly.
We want to consider each possible monomial in the $y$ variables. 
Each edge will contribute a derivative in either the real or complex direction,
so the only monomials that appear have the form
\[
\left(\frac{y^1_{i_1}}{T_1}\right)^{\ell_1} \left(\frac{y^2_{i_2}}{T_2}\right)^{\ell_2} \cdots \left(\frac{y^{k-1}_{i_{k-1}}}{T_{k-1}}\right)^{\ell_{k-1}}\left(\frac{\sum_{\alpha=1}^{k-1}y^\alpha_{i_k}}{T_k}\right)^{\ell_k}
\]
where each $\ell_j$ is either zero or one. 
Let $\vec{\ell} = (\ell_1,\ldots, \ell_k)$ denote the data that specifies this monomial, and so implicitly depends on the indices $i_j$.
(Note that $\ell_1+ \cdots + \ell_k = |S|$.)
Then we denote the associated component of the polynomial $P^S$ as
\[
\left[\frac{y}{T}\right]^{\vec{\ell}} P^S_{\vec{\ell}}\left(\frac{\wbar^1}{T_1},\ldots,\frac{\wbar^{k-1}}{T_{k-1}},\frac{\sum_{\alpha=1}^{k-1}\wbar^\alpha}{T_k}\right).
\]
We have thus obtained a further decomposition:
\begin{equation}
\label{WS decomp}
W_{\epsilon < L}^{k, (\vec{p}), S}(\Phi) = \sum_{\vec \ell} W_{\epsilon < L}^{k, (\vec{p}), S, \vec\ell}(\Phi).
\end{equation}
Much of our work will boil down to finding bounds on each $\vec\ell$ term in the $\epsilon \to 0$ limit.

Now that we have separated out the real and complex directions,
we make one final maneuver.
Using Lemma \ref{lmm: zeta} and Corollary \ref{clever diff op}, 
we note that 
\[
P^S_{\vec{\ell}}\left(\frac{\wbar^1}{T_1},\ldots,\frac{\wbar^{k-1}}{T_{k-1}},\frac{\sum_{\alpha=1}^{k-1}\wbar^\alpha}{T_k}\right)G^{(k)}_{\vec T}
= P^S_{\vec\ell} \left (\frac{\partial }{\partial w^1}-\zeta,\ldots,\frac{\partial }{\partial w^{k-1}}-\zeta,\zeta\right)G^{(k)}_{\vec T}.
\]
Having rewritten in terms of a differential operator 
\[
\sP^S_{\vec\ell} = P^S_{\vec\ell}\left (\frac{\partial }{\partial w^1}-\zeta,\ldots,\frac{\partial }{\partial w^{k-1}}-\zeta,\zeta\right)
\]
involving only holomorphic derivatives acting on $G^{(k)}_{\vec T}$,
we can apply integration by parts to obtain the following.

\begin{lmm}
\label{WSell IBP}
For any input $\Phi$,
\begin{align*}
W_{\epsilon < L}^{k, (\vec{p}), S, \vec\ell}(\Phi) = \displaystyle \int_{Y^{k}} &
\d q^k\left(\prod_{\alpha=1}^{k-1} \d^n w^\alpha\right) \int_{\vec T \in [\epsilon,L]^k} \prod_{\alpha=1}^k \d T_\alpha \\
& \displaystyle \times \left( \sP^S_{\vec\ell}\, \Phi(q^1,\ldots,q^k)\right)  \left[\frac{y}{T}\right]^{\vec\ell} G^{(k)}_{\vec T}.
\end{align*}
\end{lmm}

This result will play a role in understanding the $\epsilon$ dependence of the weights.

\subsubsection{Algebraic vanishing}

The main result of this subsection is the following. 

\begin{prp}
\label{lmm: vanish}
The weight $W_{\epsilon<L}^{k,(\vec p)}(\Phi)$ is identically zero if $k \leq m + n$. 
\end{prp}

Two different kinds of argument appear, so we obtain this proposition as a consequence of two lemmas.

%

\begin{lmm}
\label{lmm: Sbound}
For the subset $S \subset \{1, \ldots, k\}$, if its cardinality satisfies $|S| < m$ or $|S| > k-n$, 
then the weight $W_{\epsilon < L}^{k, (\vec{p}), S}(\Phi)=0$ for any input $\Phi$.
In particular, the analytic weight $W_{\epsilon<L}^{k,(\vec p)}(\Phi)$ is given by the sum of the $S$-weights just over the subsets $S$ that satisfy 
\[ 
|S| \leq k -n \quad\text{and}\quad |S|  \geq m .
\]
\end{lmm}

\begin{crl}
If $k < m + n$, the analytic weight $W_{\epsilon<L}^{k,(\vec p)}(\Phi)$ is zero. 
\end{crl}

\begin{proof}
Let $q^\alpha$ denote the ``center of mass'' coordinates on $Y^k$ introduced in Equations~\eqref{eq: coords1} and~\eqref{eq: coords2}. The product 
\begin{equation}\label{eqn:integrandtype}
\prod_{\alpha} D_\alpha \pi_\alpha^* E_{T_\alpha} = \sum_{|S| \leq k} \left(\prod_{\alpha \in S} D_{\alpha} \pi_\alpha^*E_{T_\alpha}^{\d} \prod_{\beta \in S^c} D_\beta \pi_\beta^*E^\dbar_{T_\beta}\right)
\end{equation}
can be expressed as a differential form on $Y^{k-1} \cong \RR^{m(k-1)} \times \CC^{n(k-1)}$ whose coordinates are $q^1, \ldots, q^{k-1}$,
where we have used the fact that 
\[
r^{1}-r^k = -\sum_{\alpha = 1}^{k-1} q^\alpha .
\]
Note that this expression is independent of the coordinate~$q^k$. 
The operators $D_\alpha$ do not change anti-holomorphic or topological degree of forms. 

For each $\alpha$, $E_{T_\alpha} = E_{T_\alpha}^\dbar + E_{T_{\alpha}}^\d$ is a differential form on $Y$ of total degree $n+m-1$. 
Recall from \eqref{E decomp} that the first summand $E^\dbar_{T_i}$ is a form of de Rham degree $m$ and Dolbeault degree $n-1$.
The second summand $E_{T_\alpha}^\d$ is a form of de Rham degree $m-1$ and Dolbeault degree~$n$. 

Thus, for a subset $S \subset \{1,\ldots, k\}$, 
the product $\prod_{\alpha \in S} D_{\alpha} \pi_\alpha^*E_{T_\alpha}^{\d} \prod_{\beta \in S^c} D_\beta \pi_\beta^*E^\dbar_{T_\beta}$ 
has a de Rham degree $ |S|m + (k-|S|)(m-1)$ and a Dolbeault degree $ |S|(n-1) + (k - |S|) n$.
However,  if the de Rham degree is  greater than $(k-1)m$, then the product must automatically be zero on $\RR^{(k-1)m}$.
Likewise, if the Dolbeault degree is great than $(k-1)n$,  is product is automatically zero on $\CC^{(k-1)n}$. 
Thus, we obtain the following two inequalities 
\begin{align*}
|S| n + (k-|S|)(n-1) & \leq (k-1) n \\
|S| (m-1) + (k-|S|) m & \leq (k-1) m 
\end{align*}
which simplify to the inequalities claimed. 
\end{proof}

To prove the main result of this subsection, it remains to check the edge case~$k = m+n$.

\begin{proof}[Proof of Lemma \ref{lmm: vanish}]
We have just seen that when $k < n+m$ the weight $W_{\epsilon<L}^{k,(\vec p)}(\Phi)$ vanishes.
It remains to show that when $k=m+n$ the weight still vanishes.
(The proof of this result is formally identical to the proof of an analogous result in the purely holomorphic case \cite[Lemma 3.10]{BWhol}).

Let us first set up some notation.
For $\alpha=1,\ldots, n+m-1$, define the following vector field
\[
X^\alpha = \sum_{i=1}^{n} \wbar_i^\alpha \frac{\partial}{\partial \wbar_i^\alpha} + \sum_{j=1}^m y_j^\alpha \frac{\partial}{\partial y_j^\alpha} .
\]


The expression \eqref{eqn:integrandtype} can be written as a $C^\infty(Y^n)$ multiple (which involves some Gaussian expression) of the differential form on $\cA(Y^{n-1})$:
\begin{equation}\label{eqn:edgecase}
\left(\prod_{\alpha=1}^{m+n-1} \d^n w^\alpha\right)\left(\sum_{\alpha=1}^{m+n-1} \iota_{X^\alpha} \prod_{i=1}^n \left(\sum_{\alpha = 1}^{m+n-1} \d \wbar_{i}^\alpha\right) \prod_{j=1}^m \left(\sum_{\alpha = 1}^{m+n-1} \d y_{j}^\alpha\right) \right) \prod_{\alpha=1}^{m+n-1} \left(\iota_{X^\alpha}  \prod_{i=1}^n \d \wbar_i^\alpha \prod_{j=1}^m \d y_j^\alpha\right) .
\end{equation}
It is convenient to simplify this further. 
Let
\begin{align*}
\theta & = \left(\prod_{\alpha=1}^{m+n-1} \d^n w^\alpha\right) \left(\sum_{\alpha=1}^{m+n-1} \iota_{X^\alpha} \prod_{i=1}^n \left(\sum_{\alpha = 1}^{m+n-1} \d \wbar_{i}^\alpha\right) \prod_{j=1}^m \left(\sum_{\alpha = 1}^{m+n-1} \d y_{j}^\alpha\right)\right) \\
\omega & = \prod_{\alpha=1}^{m+n-1}\left(\prod_{i=1}^n \d \wbar_i^\alpha \prod_{j=1}^m \d y_j^\alpha \right).
\end{align*}
Then \eqref{eqn:edgecase} can be written as
\begin{equation}\label{eqn:edge2}
\theta \, \big(\iota_{X^1} \cdots \iota_{X^{m+n-1}} \, \omega\big) .
\end{equation}

Recall that $Y^{m+n-1} = \RR^{m(m+n-1)} \times \CC^{n(m+n-1)}$ is real $(m+n-1)(m+2n)$ dimensional. 
The form above is of degree $(m+n-1)(m+2n)$ (so, it is not necessarily zero for simple type reasons like in the other cases). 
However, for any $\alpha = 1,\ldots, m+n-1$ we see $\theta \iota_{X^1} \cdots \Hat{\iota_{X^\alpha}} \cdots \iota_{X^{m+n-1}}\omega$ is a form of type $(m+n-1)(m+2n)+1$ so is necessarily zero on $Y^{m+n-1}$. 
Since contraction $\iota_{X}$ is a derivation for any vector field $X$ we can thus rewrite  \eqref{eqn:edge2} as
\[
\big(\iota_{X^1} \cdots \iota_{X^{m+n-1}} \theta \big) \, \omega
\]
up to an overall sign. 

Finally, for any $\alpha,\beta, i$ the operator $\iota_{X^\alpha}$ commutes with wedging with the holomorphic differential form $\d w_i^\beta$. 
Thus,
\begin{multline}
\iota_{X^1} \cdots \iota_{X^{m+n-1}} \theta \\ = \left(\prod_{\alpha=1}^{m+n-1} \d^n w^\alpha\right)\iota_{X^1} \cdots \iota_{X^{m+n-1}} \left(\sum_{\alpha=1}^{m+n-1} \iota_{X^\alpha} \prod_{i=1}^n \left(\sum_{\alpha = 1}^{m+n-1} \d \wbar_{i}^\alpha\right) \prod_{j=1}^m \left(\sum_{\alpha = 1}^{m+n-1} \d y_{j}^\alpha\right)\right) .
\end{multline}
The right hand side is zero since $\iota_X \iota_X = 0$ for any vector field $X$. 
The result follows.
\end{proof}

\subsubsection{Analytic results}

When $k > m+n$, we must examine the integrals in detail to obtain the following.

\begin{prp}
\label{prp: more than 2}
If $k> n+m$, 
then the analytic weight $W_{\epsilon<L}^{k,(\vec p)}(\Phi)$ has a well-defined $\epsilon \to 0$ limit as a distribution.
\end{prp}

To prove this proposition, we will show that for any compactly-supported input $\Phi$,
the absolute value of the weight $|W^{k,(\vec p)}_{\epsilon<L}(\Phi)|$ is bounded above by an integral whose dependence on $\epsilon$ converges as $\epsilon \to 0$.
Earlier in Section~\ref{weights and reduction}, we have already expressed this weight as a sum of terms $W_{\epsilon < L}^{k, (\vec{p}), S, \vec\ell}(\Phi)$, so we need to find bounds on each term.
The basic idea is to integrate over $Y^k$ and then examine the remaining integral over the ``scale parameters''~$T_\alpha \in [\epsilon,L]$.
This decomposition isolates the dependence on $\epsilon$,
which will always have a similar form, handled by the following lemma.



\begin{lmm}
\label{lmm: amgm}
Let $N$ and $k$ be two non-negative integers.
Then,
\[
I_{N,k}  (\epsilon , L) = \int_{(T_1,\ldots,T_k) \in [\epsilon,L]^k} \frac{\d^k T}{(T_1+\cdots+T_k)^{N}}
\]
has a limit as $\epsilon \to 0$ whenever $N < k$.
Furthermore,
\[
\lim_{L \to 0} \lim_{\epsilon \to 0} I_{N,k} (\epsilon , L)=0 
\]
when $N <k$.
\end{lmm}

We will not use this last fact in this section,
but we will need it later on in the proof of Theorem~\ref{thm: noanomaly}.)

\begin{proof}
The AM-GM inequality tells us that for any positive integer~$p$ and any collection of nonnegative real numbers $T_1,\ldots, T_p$,
\[
T_1 + T_2 + \cdots + T_p \leq p \left( T_1 T_2 \cdots T_p\right)^{1/p},
\]
and hence
\[
\frac{1}{(T_1+\cdots+T_p)^{N}} \leq \frac{1}{p^N \left( T_1 T_2 \cdots T_k\right)^{N/p}}.
\]
This inequality gives us an immediate bound for our integrals:
\[
\int_{(T_1,\ldots,T_k) \in [\epsilon,L]^k} \frac{1}{(T_1+\cdots+T_k)^{N}}\d^k T
\leq \frac{1}{k^N}\int_{(T_1,\ldots,T_k) \in [\epsilon,L]^k} \frac{1}{(T_1\cdots T_k)^{N/k}}\d^k T.
\]
The right hand side is easy to integrate because it is a product of functions each of which depends only on one $T_\alpha$.
Note that for any index~$i$, we have
\[
\int_\epsilon^L T_i^{- N/k}\d T_i  = \frac{1}{1- N/k}\left(L^{1 - N/k} - \epsilon^{1 - N/k} \right),
\]
so that there is a limit as $\epsilon \to 0$ if ${1 - N/k} > 0$.


\end{proof}

\begin{proof}[Proof of proposition]
In Section~\ref{weights and reduction} we have described the analytic data relevant for the $k$-vertex wheel for each integer $k > m+n$ and for any choice of holomorphic differential operators (arising from potential interaction terms in the THFT). 
We have already broken that analytic weight $W_{\epsilon < L}^{k, (\vec{p})}$ into a sum of weights $W_{\epsilon < L}^{k, (\vec{p}), S, \vec\ell}$ 
where $S$ runs over subsets of the edges on which we use ``topological term'' $E^\d$ of the propagator and where $\vec\ell$ records a monomial factor in the integrand arising from those edges.
Note that the number of nonzero entries in $\vec\ell$ equals $|S|$, the cardinality of~$S$.

Recall from Lemma~\ref{WSell IBP} that
\begin{align*}
W_{\epsilon < L}^{k, (\vec{p}), S, \vec\ell}(\Phi) = \displaystyle \int_{Y^{k}} &
\d q^k\left(\prod_{\alpha=1}^{k-1} \d w^\alpha\right) \int_{\vec T \in [\epsilon,L]^k} \prod_{\alpha=1}^k \d T_\alpha \\
& \displaystyle \times \left( \sP^S_{\vec\ell}\, \Phi(q^1,\ldots,q^k)\right)  \left[\frac{y}{T}\right]^{\vec\ell} G^{(k)}_{\vec T}.
\end{align*}
where
\[
\left[\frac{y}{T}\right]^{\vec{\ell}} 
= \left(\frac{y^1_{i_1}}{T_1}\right)^{\ell_1} \left(\frac{y^2_{i_2}}{T_2}\right)^{\ell_2} \cdots \left(\frac{y^{k-1}_{i_{k-1}}}{T_{k-1}}\right)^{\ell_{k-1}}\left(\frac{\sum_{\alpha=1}^{k-1}y^\alpha_{i_k}}{T_k}\right)^{\ell_k}.
\]
To bound the value of this integral, we will integrate first over $Y^k$ and examine the dependence on $\vec T$.
For any finite positive values of $T_\alpha$, that integral is manifestly finite, as $\Phi$ is compactly supported,
but the integral over $\vec T$ might be divergent,
so we need to find a convenient upper bound on the $\vec T$ integrand.

As a first step, we replace $\sP^S_{\vec\ell}\, \Phi$ by a polynomial plus a remainder, thanks to Taylor's theorem.
The integral over $Y^k$ against the polynomial term is given by computing moments of a Gaussian,
which we will examine below.
As $\vec T$ gets small (in some direction), the integral over $Y^k$ against the remainder term (which has compact support) can be split up into two pieces:
outside a small ball (possibly oblong), the Gaussian becomes exponentially small and so there is an easy bound in terms of the support of the remainder term;
and inside the ball there is an easy bound in terms of a moment of the Gaussian.

Hence, to determine the $T$-dependence, we need to compute the $T$-dependence of the Gaussian moments that appear,
tracking also the $T$-contributions from the factor $[y/T]^{\vec\ell}$ and from the differential operator $\sP^S_{\vec\ell}$.
The moments arise from integrals of the form
\[
\displaystyle \int_{Y^{k-1}} \left(\prod_{\alpha=1}^{k-1} \d w^\alpha\right) y^\mu  y^{\vec\ell} G^{(k)}_{\vec T}.
\]
where $y^\mu$ denotes some monomial function on $Y^{k-1}$.
The $y^\mu$ comes from Taylor expansion of $\sP^S_{\vec\ell}\Phi$.
Note that the moment vanishes unless $y^{\mu + \vec\ell}$ has a total degree that is even.
(Any monomial involving the complex variables $w$ also has vanishing moment.)
As explained in Section~\ref{sec on big gaussian}, 
these are sums of $T$-monomials of the form
\[
\left( \frac{T_1 T_2 \cdots T_k}{T_1 + T_2 + \cdots + T_k} \right)^{m/2+n} \frac{T^\lambda}{(T_1 + \cdots + T_k)^{|\lambda|/2}} \frac{1}{(T_1 \cdots T_k)^{m/2+n}}
\]
where the first factor arises from the determinant of the matrix $M$, where the final factor is contributed by $G$,
and where the exponent $\lambda \in \NN^k$ has some interesting features, 
thanks to Lemma~\ref{lmm std gauss}.
Notably if $y_i$ divides the monomial $y^{\mu+\vec\ell}$, then $T_i$ divides $T^\lambda$.
(These factors need not appear with the same power:
for example, the moment for $y_a y_b$ returns $T_a T_b/(\sum T_j)$ if $a \neq b$, but if $a = b$, then the moment is a sum of terms $T_a T_c/(\sum T_j)$ where $c \neq a$.) 
As $\vec\ell$ involves a product of distinct, non-repeated factors, $\lambda = \vec{\ell} + \lambda'$ for some $\lambda' \in \NN^k$.
Moreover, the total degree of $T^\lambda$ satisfies
\[
|\lambda| = |\mu| + |\vec\ell|,
\] 
where $|\mu | = m_1 + \cdots +m_k$ and $|\vec\ell|= \ell_1 + \cdots +\ell_k$ denote the total degrees of the exponents.
Thus the moment has $T$-dependence
\[
\frac{T^\lambda}{(T_1 + \cdots + T_k)^{m/2+n+|\lambda|/2}},
\]
by simplifying.
(Note that the moment vanishes unless $|\lambda|$ is even.)
The differential operator $\sP^S_{\vec\ell}$ can also contribute a factor of the form $T^\nu/(\sum_{i=1}^k T_i)^N$ where $|\nu| = N$, 
while $[y/T]^{\vec\ell}$ contributes a factor of $1/T^{\vec\ell}$.
Putting these together, we find an overall $T$-dependence
\[
\frac{T^{\lambda+\nu-\vec\ell}}{(T_1 + \cdots + T_k)^{m/2+n+N+|\lambda|/2}} = \frac{T^{\lambda'+\nu}}{(T_1 + \cdots + T_k)^{m/2+n+N+|\lambda|/2}}
\]
where $\lambda'$ is a nonnegative exponent.
We will use Lemma~\ref{lmm: amgm} to identify sufficient conditions to ensure the convergence of an integral over $\vec T$ with~$\epsilon = 0$.

There is, however, an asymmetry between the $k$th cyclic coordinate $q_k \in Y$ and the others, 
because $G$ only depends on the first $k-1$ factors of $Y^k$,
and so it is necessary to treat the cases $\ell_k = 0$ and $\ell_k = 1$ separately.
%
%

{\em Case $\ell_k = 0$:}\/ 
In this case, the contribution $T^{-\vec\ell}$ from $[y/T]^{\vec\ell}$ has no power of $T_k$ in it.

Notice that because $T_i < T_1 + \cdots + T_k$ for any $i$, we see that 
\[
\frac{T_i}{T_1+\cdots+T_k}< 1
\]
and hence 
\[
\frac{T^{\lambda'+\nu}}{(T_1 + \cdots + T_k)^{m/2+n+N+|\lambda|/2}} \leq \frac{1}{(T_1 + \cdots + T_k)^{m/2+n+ |\vec\ell|/2}}
\]
because $N = |\nu|$ and $|\lambda| = |\lambda'|+|\vec\ell|$.
We know that $|S| = |\vec\ell|$ and by Lemma~\ref{lmm: Sbound} that $m \geq |S| \leq k-n$.
Hence we see that the exponent in the denominator satisfies
\[
\frac{m+|\vec\ell|}{2}+n \leq \frac{m+n}{2} + \frac{k}{2}.
\]
Moreover, by hypothesis, $m+n < k$, so 
\[
\frac{m+|\vec\ell|}{2}+n < k.
\]
Lemma~\ref{lmm: amgm} assures us that the integral over $\vec T$ thus converges as $\epsilon$ goes to zero.

{\em Case $\ell_k = 1$:}\/ 
In this case the factor
\[
\left[\frac{y}{T}\right]^{\vec{\ell}} 
= \left(\frac{y^1_{i_1}}{T_1}\right)^{\ell_1} \left(\frac{y^2_{i_2}}{T_2}\right)^{\ell_2} \cdots \left(\frac{y^{k-1}_{i_{k-1}}}{T_{k-1}}\right)^{\ell_{k-1}}\left(\frac{\sum_{\alpha=1}^{k-1}y^\alpha_{i_k}}{T_k}\right).
\]
contributes a power of $1/T_k$ but we cannot match it with a power of $T_k$ by computing a moment of the Gaussian,
as we showed in Section~\ref{sec on big gaussian}.
On the other hand, Lemma~\ref{tau} lets us replace
\begin{align*}
W_{\epsilon < L}^{k, (\vec{p}), S, \vec\ell}(\Phi) = \displaystyle \int_{Y^{k}} &
\d q^k\left(\prod_{\alpha=1}^{k-1} \d w^\alpha\right) \int_{\vec T \in [\epsilon,L]^k} \prod_{\alpha=1}^k \d T_\alpha \\
& \displaystyle \times \left( \sP^S_{\vec\ell}\, \Phi(q^1,\ldots,q^k)\right)  \left[\frac{y}{T}\right]^{\vec\ell} G^{(k)}_{\vec T}.
\end{align*}
by
\begin{align*}
 \displaystyle \int_{Y^{k}} &
\d q^k\left(\prod_{\alpha=1}^{k-1} \d w^\alpha\right) \int_{\vec T \in [\epsilon,L]^k} \prod_{\alpha=1}^k \d T_\alpha \left( \sP^S_{\vec\ell}\, \Phi(q^1,\ldots,q^k)\right) \\
& \displaystyle \times\left(\frac{y^1_{i_1}}{T_1}\right)^{\ell_1} \left(\frac{y^2_{i_2}}{T_2}\right)^{\ell_2} \cdots \left(\frac{y^{k-1}_{i_{k-1}}}{T_{k-1}}\right)^{\ell_{k-1}}\tau^{i_k} G^{(k)}_{\vec T}
\end{align*}
where $\tau^{i_k}$ is the differential operator 
\[
\tau^{i_k} = \frac{1}{\sum_{\alpha=1}^k T_\alpha} \sum_{\alpha = 1}^{k-1} T_\alpha \frac{\partial}{\partial y_{i_k}^\alpha}.
\]
This replacement has the virtue of replacing the factor $1/T_k$ by terms of the form $T_i/\sum T_j$,
which we know how to handle effectively.

Now apply integration by parts to $\tau^{i_k}$, much as we used integration by parts with the holomorphic differential operator $\sP$.
Here, integration by parts produces two kinds of terms: one applies $\tau^{i_k}$ to $\Phi$, or one applies $\tau^{i_k}$ to the $y_i/T_i$ factors.
The first term can be handled by precisely the techniques in the case $\ell_k = 0$, as the new contributions of $T_\alpha/\sum T_j$ are bounded above by 1.
The second kind of term is also straightforward.
Applying $\tau^{i_k}$ will eliminate a factor $y_{i_k}^\alpha/T_\alpha$ from $y^{\vec\ell}$ but also will multiply by $1/\sum T_j$.
Therefore, we have reduced to studying a case like for $\ell_k=0$ but with $|\ell|\leq k-n-2$ and an extra $1/\sum_i T_i$ in the denominator.
\end{proof}

\section{The one-loop anomaly vanishes}
\label{sec:anomaly}

The main result of this section is that a THFT is anomaly-free at one-loop so long as it is not purely holomorphic.

\begin{thm}
\label{thm: noanomaly}
When $m > 0$, the topological-holomorphic gauge-fixing condition for any THFT on $\RR^m \times \CC^n$ (with or without background fields) produces an effective family which solves the quantum master equation to first order in $\hbar$. 
In other words, one obtains a QFT defined modulo $\hbar^2$. 
\end{thm}

As we recalled in Section~\ref{sec: gf}, a QFT in the BV formalism is an effective family $\{I[L]\}$ satisfying the quantum master equation. 
We have seen how a particular choice of a gauge fixing condition available for THFT's on $\RR^m \times \CC^n$ produces a one-loop finite effective family. 
To prove the theorem it suffices to show that the $L\to 0$ limit of 
\[
\hbar \Theta [L] = \big(QI[L]+\frac{1}{2}\{I[L],I[L]\}_L + \hbar\Delta_L I[L] \big) \mod \hbar^2
\]
vanishes.
The proof of this result is a little more subtle than proving the one-loop finiteness as in Theorem~\ref{thm: nodivs},
but essentially it is an elaboration of the techniques used there.
Before delving into explicit analysis, we begin by describing the structure of the one-loop diagrams and qualitative aspects of the associated integrals in the next subsection.

Throughout this section $m > 0$. 
Note that when $m = 0$, the result does not hold \cite{CLbcov1,BWhol}; 
holomorphic theories on $\CC^n$ do not necessarily admit effective families that solve the quantum master equation. 
(These are holomorphic cousins of ``chiral anomalies'' in the physics literature.)

Theorem~\ref{thm: noanomaly} has a wide degree of applicability: 
it applies to the anomalies of theories in any number of dimensions, and it applies also in the presence of background fields.
Let us, however, point out several facts that prevent the theorem from being ``too good to be true.''
First, it applies only to the one-loop anomaly of THFTs; it says nothing about higher-loop obstructions to quantization.
Second, it applies only to theories on spaces of the form $\RR^m\times \CC^n$.
Although THFTs may be formulated on more general spacetimes, such products~$M\times X$ where $M$ is an oriented manifold and $X$ is a complex manifold,
Theorem \ref{thm: noanomaly} makes no claim about the vanishing of anomalies on such spacetimes.
Indeed, there can be anomalies that depend on topological and geometric features of the global structure of the manifold.
Finally, there are symmetries of THFTs that do not furnish THFTs with background fields (cf. Example \ref{ex: nonexample}); again, Theorem \ref{thm: noanomaly} does not apply in such examples.

\subsection{The ingredients and some simplifications}

Recall that we set $Y = \RR^m \times \CC^n$.
Our proof resembles the analogous proofs in \autocite[Theorem 5.1]{GWcs} and~\autocite[Section 4]{BWhol}. 


The diagrams for the one-loop obstruction are again wheels, like the diagrams for the one-loop effective action, 
except that here there is a distinguished internal edge labeled by the scale $\epsilon$ heat kernel $K_\epsilon$ while the remaining edges are labeled by the propagator~$P_{\epsilon<L}$ (cf.  \autocite[Appendix C]{LiLi}).
(Each choice of labeled internal edge contributes to the weight of the anomaly, but the fundamental analytic aspects do not depend on which edge is labeled so we will fix a distinguished edge from hereon.)
The analytic contribution to the obstruction from a wheel diagram with $k$ vertices is thus
\[
\Theta_{\epsilon < L}^{k} (\Phi) = \sum_{|S| \leq k-1} \Theta_{\epsilon < L}^{k, S} (\Phi)
\]
where $\Phi$ is a smooth, compactly-supported section of $A$ on $(\CC^n\times \RR^m)^k$ whose value depends on the input fields---background and propagating---appearing on the external legs of the relevant diagram, and where
\begin{align}\label{eqn: oweight1}
&\Theta_{\epsilon < L}^{k, S} (\Phi) \\
&= \int_{\vec T \in [\epsilon, L]^{k-1}} \d\vec T \int_{Y^k} \left(\prod_{\alpha=1}^k \d^n z^\alpha\right)\Phi(x,z) \left(\prod_{\sigma \in S} \pi_\sigma^*E_{T_\sigma}^{\d} \prod_{\tau \notin S} \pi_\tau^*E_{T_\tau}^\dbar \right) \pi_k^*K^{an}_\epsilon\nonumber.
\end{align}
We use $\Theta_{\epsilon < L}^{k, S} (\Phi)$ to denote the term associated with the choice of~$S$, a subset of $\{1,\ldots, k-1\}$ that arises from expanding the propagators.

We sum over the subsets $S$ to get the total weight of the wheel diagram.

Note that we have assumed that there are no holomorphic derivatives appearing in the interaction vertices of our diagrams; the proof of the more general situation is not any more difficult, but it requires the introduction of bulky notation.
Since we have already shown how one approaches the more general situation in the context of Theorem \ref{thm: nodivs}, we hope that it will be clear to the reader the minor modifications necessary to make the general argument apply.

As before, certain diagrams (for small $k$ relative to $m,n$) will vanish identically for form-type reasons, while the diagrams for large $k$ will have a well-defined $\epsilon\to 0$ limit, and we will see that the $L\to 0$ limit of the $\epsilon\to 0$ limit of these diagrams is zero.
The following result is parallel to Lemma~\ref{lmm: Sbound}. 

\begin{lmm}\label{lmm: anomalyS1}
The term $\Theta_{\epsilon<L}^{k, S}(\Phi)$ vanishes unless 
\[ 
|S| \leq k -n -1\quad \text{ and } \quad |S| \geq m .
\]
In particular, $\Theta_{\epsilon<L}^{k, S}(\Phi)$ is identically zero if $k \leq m + n$. 
\end{lmm}

\begin{proof}
Let $q^\alpha$ denote the ``center of mass'' coordinates on $Y^k$ introduced in Equations \ref{eq: coords1} and \ref{eq: coords2}. Then, we see that the product 
\[
\left(\prod_{\alpha=1}^{k-1} D_\alpha \pi_\alpha^* E_{T_\alpha}\right) \pi_k K^{an}_\epsilon = \sum_{|S| \leq k-1} \left(\prod_{\sigma \in S} D_{\sigma} \pi_\sigma^*E_{T_\sigma}^{\d} \prod_{\tau \notin S} D_\tau \pi_\tau^*E_{T_\tau}\right) \pi_k^* K^{an}_\epsilon
\]
can be expressed as a differential form on $Y^{k-1} \cong \RR^{m(k-1)} \times \CC^{n(k-1)}$ whose coordinates are $q^1, \ldots, q^{k-1}$.
For each $\alpha$, $E_{T_\alpha} = E_{T_\alpha}^\dbar + E_{T_{\alpha}}^\d$ is a differential form on $Y$ of total degree $n+m-1$. 
Recall from \eqref{E decomp} that the first summand $E^\dbar_{T_i}$ is a form of  de Rham degree $m$ and Dolbeault degree $n-1$.
The second summand $E_{T_\alpha}^\d$ is a form of de Rham degree $m-1$ and Dolbeault degree $n$. 
The third term $K^{an}_{\epsilon}$ is a differential form of de Rham type $m$ and Dolbeault type~$n$. 

Just as in the proof of Lemma~\ref{lmm: Sbound}, we find the conditions
\begin{align*}
|S| n + (k-1-|S|) (n-1) + n & \leq (k-1) n \\
|S| (m-1) + (k-1-|S|) m + m & \leq (k-1) m ,
\end{align*}
which imply the statement of the Lemma.
\end{proof}

\subsection{Computing the anomaly}

By Lemma \ref{lmm: anomalyS1}, we know that the anomaly is only potentially nonzero when $k > n + m$.
The key analytic fact is the following.

\begin{lmm}
\label{lmm: anomalya}
If $k\geq m+ n+1$, 
then 
\[
\displaystyle \lim_{L \to 0} \lim_{\epsilon \to 0} \Theta_{\epsilon<L}^{k}(\Phi) = 0,
\]
i.e. the limit of the analytic weight under consideration vanishes.
\end{lmm}

\begin{proof}
We prove the claim for each weight $\Theta_{\epsilon<L}^{k,S}(\Phi)$ where we
take an arbitrary subset $S \subset \{1,\ldots, k-1\}$.
It is useful to expand out this weight in Equation~\eqref{eqn: oweight1} in much more detail.
As usual, we work in the familiar ``center of mass'' change of coordinates from Equations \eqref{eq: coords1} and \eqref{eq: coords2}, just as we did in the proof of Theorem~\ref{thm: nodivs}. 
We then write the weight itself as a sum 
\[
\Theta_{\epsilon<L}^{k,S}(\Phi) = \sum_{f,g} \Theta_{\epsilon<L}^{k,S, f,g}(\Phi)
\]
where we run over choices of a function $f: S\to \{1,\ldots, m\}$ and a function $g: S\to \{1,\ldots, n\}$. 
Here $\Theta_{\epsilon<L}^{k,S, f,g}(\Phi)$ denotes the integral
\begin{align}
\displaystyle  
\int_{\vec T \in [\epsilon, L]^{k-1}}& \d\vec T  \int_{\vec q \in Y^k} 
\d^{2n+m} q_{\Sigma} \,\prod_{\alpha=1}^k \d^n w^\alpha \, \Phi (q^1,\ldots,q^k) 
 \\
&  \times \prod_{\sigma \in S} \left(\d^n\bar {w}^\sigma\,\frac{y^\sigma_{f(\sigma)}}{T_\sigma}\prod_{i\in \{1,\ldots, m\}\backslash f(\sigma)} \d y^\sigma_i\right)  \nonumber\\
&\times\prod_{\tau\notin S}\left(\d^m y^\tau\,\left(\frac{\partial }{\partial w^\tau_{g(\tau)}}-\zeta^{g(\tau)}\right)\prod_{j\in \{1,\ldots, n\}\backslash g(\tau)}\d \bar{w}^\tau_j \right) \,G^{(k)}_{\vec T, \epsilon}\nonumber
\label{eq: gnarlyweight}
\end{align}
where $G^{(k)}_{\vec T, \epsilon}= G^{(k)}_{\vec T'}$ for the $k$-tuple $\vec T' = \{T_1,\ldots, T_{k-1},\epsilon\}$ and where, as usual, we use notation like $\d^m y^\tau$ to denote the volume form on the $\tau$-copy of $\RR^m$ in the coordinates $y^\tau_1, \ldots, y^\tau_m$.
Note that $q_\Sigma$ denotes a copy of $\RR^m \times \CC^n$ with coordinate the sum $q^1 + \cdots + q^{k-1}$, so $\d^{m + 2n} q_\Sigma$ denotes the volume form on it.
(In the original $r$ coordinates, $q_\Sigma = r_k - r_1$. See Section~\ref{sec:com}.)

To shorten this expression, we let $P_{S,g}(\vec T')$ denote the $\vec T'$-dependent holomorphic differential operator 
\begin{equation}
\prod_{\sigma\in S} \prod_{j=1}^n \left(\d \bar{w}^{\sigma}_j\right) \prod_{\tau\notin S}\left(\left( \frac{\del}{\del w^\tau_{g(\tau)}}-\zeta_{g(\tau)}\right)\prod_{\substack{j\in \{1,\ldots, n\}\backslash g(\tau)}}\d \bar{w}^\tau_j \right).
\end{equation}

The operator $P_{S,g}$ depends on $\vec T$ through the definition of~$\zeta$, 
which is defined in Lemma~\ref{lmm: zeta}.
We then have
\begin{align*}
\displaystyle  
\Theta_{\epsilon<L}^{k,S, f,g}(\Phi) =
\int_{\vec T \in [\epsilon, L]^{k-1}} &\d\vec T  \int_{\vec q \in Y^k} 
\d^{2n+m} q_\Sigma \,\prod_{\alpha=1}^k \d^n w^\alpha \, \Phi (q^1,\ldots,q^k)  \nonumber
 \\
& \times \prod_{\sigma \in S} \left(\frac{y^\sigma_{f(\sigma)}}{T_\sigma}\prod_{\substack {i\in \{1,\ldots, m\}\\i\neq f(\sigma)}} \d y^\sigma_i \right) 
P_{S,g}(\vec T) \,G^{(k)}_{\vec T, \epsilon} \nonumber
\end{align*}
(Later, we will use integration by parts to move $P_{S,g}$ onto the $\Phi$ term.)

We now sketch the essential idea of the proof, 
before wading into explicit integrals.
Recall that the Gaussian expression $G^{(k)}_{\vec T,\epsilon} = G_{\vec{T},\epsilon}^{(k)}(q^1,\ldots,q^{k-1})$ only depends on the center of mass coordinates $q^1,\ldots, q^{k-1}$.
We will perform the $(k-1)(2n + 1)$-dimensional Gaussian integral over $(q^1,\ldots, q^{k-1}) \in Y^{k-1} = (\RR^m \times \CC^n)^{(k-1)}$. 
This Gaussian integral can be written as a sum over the Taylor components of the compactly supported functions~$\Phi$. 
Our basic strategy is to find bounds on the moments of these Taylor components and on the integrals of the Taylor remainders.

The argument now forks into two cases: the one in which $|S|$ is odd and the one in which $|S|$ is even. 

{\em Case $|S| = 2N+1$ odd}\/:
Without loss of generality, we can assume that $S = \{1,\ldots, 2N+1\} \subset \{1,\ldots, k-1\}$; otherwise we can just permute the indices.

By Taylor's Theorem, and using the explicit integral form of the remainder, we may write:
\[
\Phi(y^1,z^1,\ldots, y^k, z^k) = \Phi(0,z^1,\ldots, 0,z^{k-1},y^k,z^k) +\sum_{\ell=1}^{m}\sum_{\delta=1}^{k-1}y^\delta_\ell R_{\delta,\ell}(q^1,\ldots, q^k).
\]
For each $\delta, \ell$, the remainder $R_{\delta, \ell}$ is uniformly bounded on $(\RR^m\times \CC^n)^k$ by a number that depends only on the supremum norms of $\Phi$ and its derivatives.
The Gaussian integral involving the constant term of the Taylor expansion vanishes, 
since it is an odd moment of a Gaussian density.
The remainder term in the Taylor expansion of $\Phi$ yields the weight:
\begin{align}\label{eq: gnarlyweight2}
\displaystyle  
\int_{\vec T \in [\epsilon, L]^{k-1}} \d\vec T  \int_{\vec q \in Y^k} 
& \d^{2n+m} q_\Sigma \, \prod_{\alpha=1}^k \d^n w^\alpha\, \left( \sum_{\substack{\ell=1\\\delta=1}}^{m,k-1}R_{\delta,\ell}(q^1,\ldots, q^k) y^\delta_\ell\right) \nonumber\\
&  \times \prod_{\sigma \in S} \left(\frac{y^\sigma_{f(\sigma)}}{T_\sigma}\prod_{\substack {i\in \{1,\ldots, m\}\\i\neq f(\sigma)}} \d y^\sigma_i \right) 
P_{S,g}(\vec T) \,G^{(k)}_{\vec T, \epsilon} \nonumber
\end{align}
We integrate by parts to have $P_{S,g}$ act on the remainder term $R_{\delta,\ell}$:
\begin{align*}
\displaystyle  
\int_{\vec T \in [\epsilon, L]^{k-1}} \d\vec T  \int_{\vec q \in Y^k} 
&\d^{2n+m} q_\Sigma \, \prod_{\alpha=1}^k \d^n w^\alpha\, \left( \sum_{\substack{\ell=1\\\delta=1}}^{m,k-1}\left(-P_{S,g}R_{\delta,\ell}y^\delta_\ell\right) \right)\nonumber
 \\
&  \times \left(\prod_{\sigma \in S} \frac{y^\sigma_{f(\sigma)}}{T_\sigma}\prod_{\substack {i\in \{1,\ldots, m\}\\i\neq f(\sigma)}} \d y^\sigma_i\right)\left(\prod_{\substack{\tau\notin S\\i\in \{1,\ldots, m\}}}\d y^\tau_i\right) \,G^{(k)}_{\vec T, \epsilon}
\end{align*}
Now, the absolute value $|P_{S,g}\frac{\del}{\del y^\delta_\ell} R_{\delta, T}|$, as a function of the $q$ and $T$ variables, is bounded above by an expression depending in a simple way on the the $C^M$-norm of~$\Phi$ for $M$ sufficiently large:
\[
\left|P_{S,g} R_{\delta,\ell} \right|(q,T) \leq C_{S,g} |\Phi|_{C^M}.
\]
Taking this into account, and performing the Gaussian integral over the $q$-coordinates, 
we find that the weight for the Taylor remainder term in $\Phi$ is bounded by
\begin{align}\label{eq: taylorexpandedTintegral}
\sum_{\sigma=1}^{2N+1} C'_{S,g,\delta}&|\Phi|_{C^M}\int_{\vec T \in [\epsilon, L]^{k-1}} \d\vec T  \frac{(T_1 \cdots T_{k-1} \epsilon)^{n + m/2}}{(T_1 + \cdots + T_{k-1} + \epsilon)^{n+m/2}}\\
& \times\frac{1}{(T_1 \cdots T_{k-1} \epsilon)^{n + m/2}} \frac{1}{T_1 \cdots T_{2N+1}} \frac{T_1 \cdots \Hat{T}_\sigma \cdots T_{2N+1}}{(T_1 + \cdots + T_{k-1} + \epsilon)^N} (M^{-1}_{\vec T, \epsilon})_{\sigma\delta}\nonumber,
\end{align}
where $C'_{S,g,\delta}$ does not depend on $\Phi$.
Here, $M^{-1}_{\vec T, \epsilon}$ is the inverse of the matrix $M_{\vec T'}$---defined in Equation~\eqref{eq: Mdefn}---associated to the $k$-tuple $\vec T'=\{T_1,\ldots, T_{k-1}, \epsilon\}$. 
Each factor  in the integrand has a simple explanation from Gaussian integration: 
the first factor arises from the computation of the determinant of $M_{\vec T,\epsilon}$; 
the second factor arises from the normalization $1/(T)^{n+m/2}$, which appears in each propagator and heat kernel; 
the third factor arises from the factors $y^\sigma_{f(\sigma)}/T_\sigma$ appearing in the integrands under consideration; 
and the last factor arises from applying Wick's theorem to compute the relevant Gaussian moments in terms of matrix entries of~$M_{\vec T, \epsilon}^{-1}$.

Recall from Section~\ref{sec on big gaussian} that the matrix element $M_{\sigma\delta}^{-1} (\vec T, \epsilon)$ is bounded above by $T_{\sigma}$ for all $\delta$. 
We can thus bound the integral \eqref{eq: taylorexpandedTintegral} by
\[
C''_{S,g,\delta } |\Phi|_{C^M}\int_{\vec T \in [\epsilon, L]^{k-1}} \d\vec T\frac{1}{(T_1 + \cdots + T_{k-1} + \epsilon)^{N+n+m/2}}
\]
Our goal is to show that $\epsilon \to 0, L \to 0$ limit of the right-hand side is zero. 
Notice that for $\epsilon > 0$, the integral above is bounded by
\[
C''_{S,g}|\Phi|_{C^M}\int_{(T_1,\ldots,T_{k-1}) \in [\epsilon,L]^{k-1}} \frac{\d^k T}{(T_1+\cdots+ T_{k-1})^{n+m/2+N}} .
\]
By Lemma~\ref{lmm: anomalyS1}, we may assume $2N \leq k-n-2$, 
whence Lemma~\ref{lmm: amgm} shows that the $\epsilon \to 0$, $L \to 0$ limit of this quantity is zero if $k > n+m$, which does indeed hold by the assumptions of the lemma.

{\em Case $|S| = 2N$ even}\/:
Without loss of generality we can assume that $S = \{1,\ldots, 2N\} \subset \{1,\ldots, k-1\}$;
otherwise we just permute the indices.
We break up the argument into two subcases, depending on the size of~$k$.

{\em Subcase $k>n+m+1$}\/:
As in the case with $|S|$ odd, we may integrate by parts to bring the holomorphic derivatives to act on $\Phi$.
As in the case $|S|$ odd, we can bound $|P_{S,g}\Phi|$ uniformly in terms of the $C^M$ norm of $\Phi$ for $M$ sufficiently large.
Upon performing the Gaussian integral, therefore, we find it is bounded above by
\begin{align*}
C_{S,g}|\Phi|_{C^M} &\int_{\vec T \in [\epsilon, L]^{k-1}} \d\vec T  
\frac{(T_1 \cdots T_{k-1} \epsilon)^{n + m/2}}{(T_1 + \cdots + T_{k-1} + \epsilon)^{n+m/2}} \frac{1}{(T_1 \cdots T_{k-1} \epsilon)^{n + m/2}} \frac{1}{T_1 \cdots T_{2N}} \frac{T_1 \cdots T_{2N}}{(T_1 + \cdots + T_{k-1} + \epsilon)^N} \\
&=C_{S,g}|\Phi|_{C^M}\int_{\vec T\in [\epsilon,L]^{k-1}}\d \vec T\, \frac{1}{(T_1+\ldots+T_{k-1}+\epsilon)^{n+m/2+N} }\\
&\leq C_{S,g}|\Phi|_{C^M}\int_{\vec T\in [\epsilon,L]^{k-1}}\d \vec T\, \frac{1}{(T_1+\ldots+T_{k-1})^{n+m/2+N} }.
\end{align*}
We now want to show that the $\epsilon \to 0, L \to 0$ limit of the right-hand side is zero. 
Lemma~\ref{lmm: anomalyS1} shows that we may assume $2N \leq k-n-1$, 
so Lemma~\ref{lmm: amgm} guarantees that the $\epsilon \to 0$, $L \to 0$ limit of this quantity is zero provided~$k-1 > n+m$, which is our hypothesis for the subcase under consideration. 

{\em Subcase $k = n+m+1$}\/:
Under this hypothesis on $k$, 
the inequalities of Lemma \ref{lmm: anomalyS1} imply that $|S|=2N=m$.
In particular, the lemma is proved for $m$ odd.

Below, we will use the equality $|S|=m$ repeatedly.
To study this case, we will need to perform a first-order Taylor expansion of $\Phi$ in the variables $\{y^1,\ldots, y^{k-1}\}$, keeping for consideration also the remainder term.
We will need to argue separately that the zeroth-order, first-order, and remainder summands of this Taylor expansion contribute zero in the $L\to 0, \epsilon \to 0$ limit.
The first-order term in the Taylor expansion gives zero after the Gaussian integration, because the Gaussian integration computes an odd moment of the Gaussian distribution.
As for the remainder term, by similar arguments to those that produced the term in Equation \eqref{eq: taylorexpandedTintegral}, one can show that the norm of the weight associated to the remainder term of the Taylor expansion is bounded by
\begin{align*}
\sum_{\delta,\eta}&C_{\delta\eta}|\Phi|_{C^M}\int_{\vec T\in [\epsilon,L]^{n+m}} \d \vec T\,\frac{1}{T_1\cdots T_{2N}}\frac{T_1\cdots T_{2N}}{(T_1+\cdots+T_{n+m}+\epsilon)^{n+m/2+N}} (M^{-1}_{\vec T, \epsilon})_{\delta \eta} \\+
&\sum_{\delta, \eta}C'_{\delta\eta}|\Phi|_{C^M}\int_{\vec T\in [\epsilon,L]^{n+m}}\d \vec T \frac{1}{T_1\cdots T_{2N}}\frac{T_1\cdots \hat T_\sigma \cdots \hat T_{\tau}\cdots T_{2N}}{(T_1+\cdots+T_{n+m}+\epsilon)^{n+m/2+N-1}}(M^{-1}_{\vec T, \epsilon})_{\sigma \delta} (M^{-1}_{\vec T, \epsilon})_{\tau \eta} .
\end{align*}
Here, $M$ is a sufficiently large integer, while the $C_{\delta\eta}$ and $C'_{\delta\eta}$ are constants that are independent of $\Phi$ (and not necessarily equal to the quantities denoted by the same symbol for other parts of the proof of this lemma).
The two sums arise from an application of Wick's lemma to compute the moment of $y^1\cdots y^{2N} y^\delta y^\eta$, with the first $2N$ factors arising from propagators, and the last two factors arising from the remainder term of the Taylor expansion.
In particular, the first sum arises from pairings of the $2N+2$ variables that pair $y^\delta$ with $y^\eta$;
the second sum arises from pairings in which $y^\delta$ and $y^\eta$ are not paired with each other.

As before, we use the fact that $(M^{-1}_{\vec T, \epsilon})_{\sigma \delta}$ is bounded by $T_\sigma$, and $(M^{-1}_{\vec T, \epsilon})_{\delta \eta}/(T_1+\cdots+T_{n+m}+\epsilon)$ is bounded by $1$.
It therefore suffices to show that the $L\to 0, \epsilon \to 0$ limit of the following integral vanishes:
\begin{equation}
\int_{\vec T\in [\epsilon, L]^{k-1}]} \frac{\d \vec T}{(T_1+\cdots+T_{n+m}+\epsilon)^{n+m/2+N-1}}\leq \int_{\vec T\in [\epsilon, L]^{k-1}]}\frac{\d \vec T}{(T_1+\cdots+T_{n+m})^{n+m/2+N-1}}.
\end{equation}
The vanishing of this $L\to 0,\epsilon \to 0$ limit is implied by the equality $N=m/2$ and Lemma~\ref{lmm: amgm}, as in many preceding arguments.

Finally, we need to consider the weight associated to the zeroth-order term in the Taylor expansion of~$\Phi$:
\begin{multline}
\label{eq: gnarlyweight3}
\displaystyle  
\int_{\vec T \in [\epsilon, L]^{k-1}} \d\vec T  \int_{\vec q \in Y^k} 
\d^{m+2n} q_\Sigma \, \prod_{\alpha=1}^k \d^n w^\alpha\, \Phi (0,z^1,\ldots,0,z^{k-1},y^k,z^k)
 \\
 \times \left(\prod_{\sigma \in S} \frac{y^\sigma_{f(\sigma)}}{T_\sigma}\prod_{\substack {i\in \{1,\ldots, m\}\\i\neq f(\sigma)}} \d y^\sigma_i\right) P_{S,g} \,G^{(k)}_{\vec T, \epsilon}.
\end{multline}
This integral vanishes for all $\epsilon, L$ by a form-type argument (a different form-type argument from that of Lemma \ref{lmm: anomalyS1}).
If $f: S\to \{1,\ldots, m\}$ is a bijection, then the Gaussian integral is zero because $y^\sigma_i$ and $y^\tau_{i'}$ are uncorrelated in the probability distribution determined by $G^{(k)}_{\vec T,\epsilon}$ when $i\neq i'$.
If $f$ is not a bijection, there is an $i_0$ that is not in the image of $f$ (this is the part of the proof that uses $m>0$).
Hence, the quantity
\[
\prod_{\substack{\sigma\in S\\i\in \{1,\ldots, m\}\\i\neq f(\sigma)}}\d y^\sigma_i
\]
contains a factor
\[
\prod_{\sigma\in S} \d y^\sigma_{i_0}.
\]
The weight in Equation~\eqref{eq: gnarlyweight3}, therefore, carries a factor
\[
\prod_{\sigma\in S} \d y^\sigma_{i_0}\prod_{\tau\notin S}\d y^\tau_{i_0} \sum_{\gamma=1}^{k-1}\d y^\gamma_{i_0}, 
\]
which is manifestly zero.
\end{proof}

\section{Examples and applications} 

We turn to some examples and applications of our main results.
We revisit, of course, the examples from the introduction, mostly discussing how our results relate to prior work.
We also examine the first example of a THFT with background fields, which offers the THFT analogue of the Adler--Bardeen--Jackiw (and also Konishi) anomalies.
Finally, we discuss the rotational symmetry of two-dimensional topological BF theory, viewed as a THFT,
and how it shows some limits of our results.

Many more applications are possible, but the most interesting demand that one delve more deeply into the theory under consideration.
Here we have simply focused on examples that quickly exhibit our results and might show a reader already familiar with QFT and anomalies how they are useful.

\subsection{Charged matter is anomaly free}

In this section, we revisit the context of Section~\ref{sec: chgdmatter}.
Recall that there we used the data of a Lie algebra $\fg$ and a $\fg$-module $P$ to define a THFT with background fields on $\RR^m\times \CC^n$.
The background fields in this case correspond to the Lie algebra~$\cG = \cA \otimes \fg$.

Theorem \ref{thm: nodivs} shows that this THFT with background fields has no one-loop anomalies as long as $m\geq 1$.
Moreover, because of the combinatorics of this particular set-up (each vertex can contribute only two half-edges to the creation of internal edges),
the full anomaly is contained in its one-loop information.
Hence, we have the following Corollary of Theorem~\ref{thm: nodivs}:

\begin{crl}
When $m\geq 1$, the charged matter THFT with background fields is anomaly-free on~$\RR^m\times \CC^n$.
\end{crl}

Physically, theories with matter consisting of fermions that obey a classical symmetry are potentially plagued by so-called ``chiral anomalies'' at the quantum level. 
Perhaps the most famous instance of this is the Adler--Bell--Jackiw anomaly in quantum electrodynamics~\autocite{Adler, BJ}. 
Another instance is in $\cN=1$ supersymmetry, where the analogous anomaly to quantizing the action of a supersymmetric background gauge field is called the Konishi anomaly~\autocite{Konishi}. 

The theories of charged matter we consider here are THFT analogs of more familiar theories of matter.
For instance, in dimensions $2 \leq d \leq 6$, the THFTs under consideration are equivalent to twists of certain theories of supersymmetric matter \autocite{ESW}.
This result implies that as soon as there is a single topological direction, no such chiral anomalies to quantizing a background gauge field are present to first-order in~$\hbar$. 

Taking $\cG = \Omega^\bu \otimes \fg$, this example is also closely related to generalizations of the axial symmetry (which involves $G = U(1)$);
the above Corollary therefore implies that there is no anomaly for these versions of the axial symmetry.
(See \autocite{rabaxial} for a treatment of this anomaly on closed spacetimes in the formalism used in this paper.)

\subsection{Chern--Simons theories in various dimensions}
\label{sec: CS revisited}

We revisit the mixed Chern--Simons theories from the introduction.
Note that ordinary Chern--Simons theory (and a natural class of related THFTs) is treated by our  methods in~\autocite{GWcs}.

\subsubsection{Four-dimensional Chern--Simons theory}

Recall from Section~\ref{sec:4dcs} that four-dimensional Chern--Simons theory is a THFT on $\RR^2 \times \CC$. 
Our results thus imply the following.

\begin{prp}
Four-dimensional Chern--Simons theory on $\RR^2 \times \CC$ admits a finite one-loop quantization.
\end{prp}

Let us describe how our result fits into recent work on this theory.
One of the main results of \autocite{CostelloYangian}, which initiated the study of this theory,
is that a unique quantization exists {\it to all orders in $\hbar$} for four-dimensional Chern--Simons theory.
(See Section 16 of that paper where he computes the relevant obstruction-deformation complex and shows that the obstruction space vanishes.)
His arguments also work over a large class of four-manifolds, because they are cohomological in nature. 

The primary advantage of our approach is that we give an explicit construction of the one-loop quantization. 
Using the holomorphic-topological gauge on $\RR^2 \times \CC$, 
we construct an explicit effective family $I[L]$ to first-order in $\hbar$ that satisfies the QME
\[
QI[L]+\frac{1}{2}\{I[L],I[L]\}_L + \hbar\Delta_L I[L]=0 
\]
modulo $\hbar^2$. 
To extend to higher powers of $\hbar$, one has to analyze more complicated integrals;
it would likely use bounds like those we have used here.

Subsequent to \autocite{CostelloYangian}, the papers \autocite{CWY1, CWY2} examined the problem of quantizing four-dimensional Chern--Simons theory in the presence of line operators, with detailed computations to first order in $\hbar$.
They use the same gauge, but much of their work leverages Lie-theoretic information.
Our paper explores how much can be seen just from the analytic structure of THFTs and how much extends beyond this four-dimensional Chern--Simons theory.
Interestingly, they find that to {\em second order} in $\hbar$, there {\em is} an anomaly to quantizing four-dimensional Chern--Simons theory in the presence of line operators. 
Their paper is a {\em tour de force} and contains a wealth of insights that we hope get absorbed into mathematics,
as they suggest rich connections between representation theory, analysis, and higher categories.
As a step in that direction, we (with Benoit Vicedo) are exploring how to lay careful foundations in the BV/renormalization package of \autocite{CosBook} for constructing the line and surface defects explored by Costello, Witten, and Yamazaki;
our efforts indicate that the results in this paper carry over, modulo some nontrivial issues about encoding defects properly with BV theories.

We now wish to flag an interesting issue that arises from the fact that we can view four-dimensional Chern--Simons theory as a holomorphic theory on $\CC^2$
(just as for any THFT on $\RR^2 \times \CC$). 
In this view, we describe the fields as 
\begin{align*}
A & \in \Omega^{0,\bu}(\CC^2 , \fg) [1] \\
B & \in \Omega^{0, \bu}(\CC^2, \fg) 
\end{align*}
where the action reads
\[
\int_{\CC^2} \d^2 z \, \kappa (B \wedge \dbar A) + \int_{\CC^2} \d^2 z \, \kappa\left(A \wedge \frac{\partial}{\partial z_1} A\right) + \frac12 \int_{\CC^2} \d^2 z \, \kappa (B \wedge [A,A]) .
\]
This action is completely equivalent to the description of four-dimensional Chern--Simons theory where $z_1$ is a holomorphic coordinate on~$\RR^2$. 
 
Note that while we have shown that THFTs on $\RR^2 \times \CC$ admit finite one-loop quantizations, 
our result does {\it not} apply to holomorphic theories on $\CC^2$. 
On the other hand, the work in \autocite{BWhol} produces an effective family for any holomorphic theory and provides a formula for the potential one-loop anomaly. 
Applied to the case at hand, the {\em holomorphic} gauge on $\CC^2$ provides a one-loop finite effective family $\{I^{hol} [L]\}$, 
and by \autocite[Proposition 4.4]{BWhol} the one-loop anomaly is the following. 

\begin{prp}
The local obstruction for the effective family $\{I^{hol}[L]\}$ to satisfy the QME to first order in $\hbar$~is 
\[
\Theta (A) = \int_{\CC^2} {\rm Tr}_{\fg^{\rm ad}}(A \wedge \partial A \wedge \partial A). 
\]
\end{prp}

%

To summarize, on one hand the topological-holomorphic gauge on $\RR^2 \times \CC$ produces a one-loop effective family $\{I[L]\}$ satisfying the QME to first order in $\hbar$,
but on the other hand, the holomorphic gauge on $\CC^2$ produces a one-loop effective family $\{I^{hol}[L]\}$ that {\it does} have an obstruction to satisfying the QME to first-order in~$\hbar$ for the exact same theory. 
Costello's result on the uniqueness of quantization guarantees that these two effective families yield equivalent quantizations. 
How can this be possible if it appears that the holomorphic gauge does {\em not} even provide a quantization? 

This issue is resolved by noticing that the local obstruction $\Theta (A)$ is {\em cohomologically trivial}. 
Indeed, consider the local functional
\[
J(A,B) = \int_{\CC^2} \d z_1 \, {\rm Tr}_{\fg^{\rm ad}} (B \wedge A \wedge \partial A) .
\]
The BV bracket of the kinetic piece of the action $\int \d^2 z A \partial_{z_1} A$ with $J$ is precisely the anomaly cocycle $\Theta$. 
The BV brackets of all other terms in the action with $J$ are zero, so one observes that $\{S, J\} = \Theta$, hence $\Theta$ is trivializable. 

\subsubsection{Five-dimensional Chern--Simons theory}

In Section \ref{sec:5dcs} we cast five-dimensional Chern--Simons theory as a THFT on $\RR \times \CC^2$. 
Our results thus imply the following.

\begin{prp}
Five-dimensional Chern--Simons theory on $\RR \times \CC^2$ admits a finite one-loop quantization.
\end{prp}

We compare this to \autocite{CostelloM5}, where Costello provides a cohomological argument that a certain ``non-commutative'' deformation of five-dimensional Chern--Simons theory admits an essentially unique quantization. 
He furthermore details how the quantization is related to a two-variable version of the Yangian quantum group. 
It would be interesting to use the holomorphic-topological gauge to provide a direct computation of this two-variable quantum group to first order in the perturbation parameter. 

\subsection{Four-dimensional gauge theory}
\label{sec: susytwist revisited}



We now turn to some examples involving four-dimensional supersymmetric gauge theory. 
Our comments here do not address the most interesting question:
how do our results relate to the role of holomorphy and of useful cancellations in the perturbative quantization of supersymmetric theories?

In Section \ref{sec:twistedsusybv}, motivated by four-dimensional $\cN=2$ supersymmetry, we introduced a two-parameter family of holomorphic gauge theories on $\CC^2$
whose fields are
\begin{align*}
A + \ep A'& \in \Pi \Omega^{0,\bu}(\CC^2 , \fg[\ep]) \\
B + \ep B' & \in \Pi \Omega^{2,\bu}(\CC^2, \fg^* [\ep]) .
\end{align*}
This BV theory is $\ZZ/2$-graded; $\Pi(-)$ denotes parity shift, and $\ep$ is an odd parameter. 
The action reads
\begin{align*}
S_{u,v}  = \int_{\CC^2} &B \wedge \dbar A' + \int_{\CC^2} B' \wedge \dbar A + \frac12 \int_{\CC^2}  B' \wedge [A,A] + \int_{\CC^2} B \wedge [A, A']  \\ 
& + u \int_{\CC^2} B \wedge \frac{\partial}{\partial w} A + v \int_{\CC^2} B' \wedge A' .
\end{align*}

For $u \ne 0$, this theory is equivalent to a $v$-dependent deformation (by a holomorphic differential operator) of topological--holomorphic BF theory on $\RR^2 \times \CC$ for the Lie algebra $\fg$.
Thus, for $u \ne 0$ it follows immediately from Theorem \ref{thm: noanomaly} that the one-loop anomaly vanishes. 

By the following proposition, we see that even at $u = 0$ the anomaly vanishes. 
In fact we will prove a stronger result, that the {\em family} of holomorphic theories has no anomaly. 
This argument has two appealing features: (1) it is uniform in the parameters $u,v$ and (2) it is completely {\em algebraic} in that it does not rely on the explicit form of the propagator or heat kernel.
This argument can be contrasted with the argument for the vanishing of the anomaly as a THFT on $\RR^2 \times \CC$ that followed from the {\em analytic} properties of the topological--holomorphic propagator. 

\begin{prp}
This family of holomorphic theories has vanishing one-loop anomaly. 
\end{prp}

\begin{proof}
By Theorem \autocite{BWhol}, the one-loop anomaly for a holomorphic theory on $\CC^2$ is given as a sum over wheels with three vertices with internal edges labeled by propagators and heat kernels. 
The analytic factors play no role in this proof, so we ignore them. 

Schematically, the propagator is sum of two terms:
\[
\begin{tikzpicture}

\node at (-3,0) {$P \;\; = $};
\draw[fermion] (-2,0) -- (0,0);
\node at (0.5,0) {$+$};
\draw[fermion] (1,0) -- (3,0);
\node at (-1.9,0.25) {$A$};
\node at (-0.1,0.25) {$B'$};
\node at (1.1,0.25) {$A'$};
\node at (2.9,0.25) {$B$};

\end{tikzpicture}
\]

The $u,v$-dependent interaction of the holomorphic theory is
\[
\frac12 \int_{\CC^2}  B' \wedge [A,A] + \int_{\CC^2} B \wedge [A, A'] + u \int_{\CC^2} B \wedge \frac{\partial}{\partial w} A + v \int_{\CC^2} B' \wedge A' .
\]
As the anomaly is computed by a graph with three vertices, the form of the interaction implies that the anomaly has components that are zero, one, two and three-linear in the gauge field $A$. 

Consider the two quadratic terms in the interaction $u \int_{\CC^2} B \wedge \frac{\partial}{\partial w} A$ and $v \int_{\CC^2} B' \wedge A'$. 
Through the BV bracket, these induce differential operators of order one and zero, respectively, acting on the fields. 
Since the gauge fixing operator $Q^{GF}$ commutes with each of these operators, we see that all graphs involving at least one bivalent vertex is identically zero. 

Thus, for the family of holomorphic theories on $\CC^2$ the only potential anomaly is three-linear in the gauge field $A$. 
There are two wheel graphs that contribute, they are drawn in Figure~\ref{fig:anomaly_kw}.

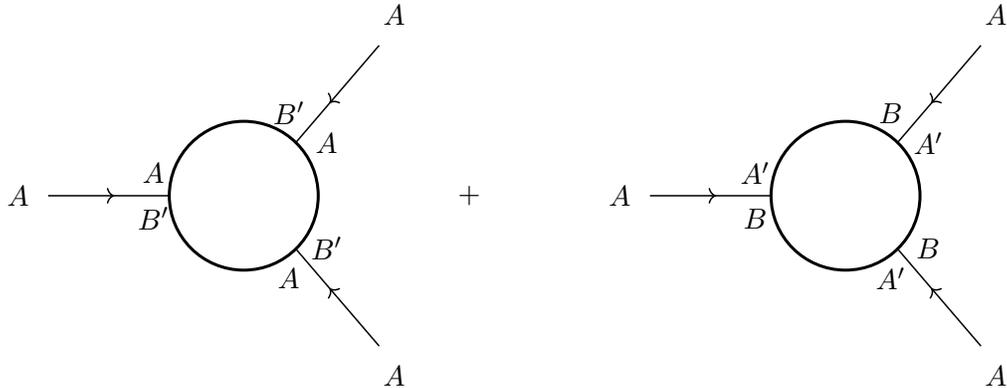
\begin{figure}
\begin{tikzpicture}[line width=.2mm, scale = 2]

		\draw[fermion] (-0.6, 1)--(-1.2, 0.3);
		\draw[fermion] (-0.6, -1)--(-1.2, -0.3);
		\draw[fermion] (-2.8, 0)--(-2, 0);
		\draw[fill=black] (-1.5,0) circle (.5cm);
		\draw[fill=white] (-1.5,0) circle (.49cm);
		\draw (-0.5, 1.2) node {$A$}; 
		\draw (-0.5, -1.2) node {$A$}; 
		\draw (-3, 0) node {$A$}; 
		
		\node at (-2.1, 0.15) {$A$};
		\node at (-1.2, 0.55) {$B'$};
		\node at (-0.95, 0.35) {$A$};
		\node at (-0.95, -0.35) {$B'$};
		\node at (-1.2, -0.55) {$A$};
		\node at (-2.1, -0.15) {$B'$};

		\draw[fermion] (3.4, 1)--(2.8, 0.3);
		\draw[fermion] (3.4, -1)--(2.8, -0.3);
		\draw[fermion] (1.2,0)--(2, 0);
		\draw (3.5, 1.2) node {$A$}; 
		\draw (3.5, -1.2) node {$A$}; 
		\draw[fill=black] (2.5,0) circle (.5cm);
		\draw[fill=white] (2.5,0) circle (.49cm);
		\draw (1, 0) node {$A$}; 
		
		\node at (1.9, 0.15) {$A'$};
		\node at (2.8, 0.55) {$B$};
		\node at (3.05, 0.35) {$A'$};
		\node at (3.05, -0.35) {$B$};
		\node at (2.8, -0.55) {$A'$};
		\node at (1.9, -0.15) {$B$};
		
		\node at (0,0) {$+$};

\clip (0,0) circle (.3cm);
\end{tikzpicture}
\caption{The anomaly for the purely holomorphic theory on $\CC^2$.}
\label{fig:anomaly_kw}
\end{figure}
It is a direct calculation to show that these two wheels contribute the same overall analytic factor, but with opposite signs.
Therefore, the one-loop anomaly vanishes.
(See, for instance, \autocite[Section 7.3]{SWsuco} for a similar argument.)
\end{proof}

\subsection{A subtlety about anomalies and background fields}

Recall that a THFT with background fields is described by a local Lie algebra of the form $\cG \oplus \cL$ where $\cL$ is a non-degenerate THFT but the background fields $\cG$ need {\em not} be described by some topological-holomorphic Lie algebra. 
For a THFT with background fields on $\RR^2$, Theorem~\ref{thm: noanomaly} implies the vanishing of the one-loop anomaly.

Recall now Section~\ref{sec: bfexamples},
where we discussed 2-dimensional topological BF theory and various ways that vector fields and holomorphic vector fields acted as symmetries.
Although 2-dimensional topological BF theory is topological, 
in Example~\ref{ex: semiexample}, 
we saw that we had to view the theory as holomorphic to include holomorphic vector fields as background fields.
Hence Theorem~\ref{thm: noanomaly} does {\em not} imply the vanishing of the one-loop anomaly for this theory with background fields;
we will now compute it explicitly and see it is nonzero.
It is, in fact, an avatar of a well-known anomaly for 2-dimensional topological BF theory and hence also familiar to those who work with the topological B-model or the Poisson $\sigma$-model. 
(The interested reader should examine \autocite{LMYonBF}, a recent paper that treats this BF theory in great depth.)

The background fields are given by the dg Lie algebra $\HH {\rm Vect}^{1,0}(\CC)$, which is
\[
\begin{tikzcd} 
{\rm Vect}^{1,0}(\CC)[1] \ar[r, "{\rm id}"] & {\rm Vect}^{1,0}(\CC), 
\end{tikzcd}
\]
where ${\rm Vect}^{1,0}(\CC)$ is the Lie algebra of smooth vector fields of type $(1,0)$ on $\CC$. 
We will denote an explicit element of this dg Lie algebra as~$(Z',Z)$, where~$Z'$ is a vector field placed in degree~$-1$ and~$Z$ is a vector field placed in degree~$0$.
If $Z' = f (z,\zbar) \partial_z$ is such a vector field, 
we denote its Jacobian by $J(Z') = \partial_z f (z,\zbar)$, which lives in~$C^\infty(\CC)$.  

\begin{prp}\label{prp:frame}
There is a one-loop obstruction to the quantization of topological BF theory with background fields given by $\HH {\rm Vect}^{1,0} (\CC)$. 
The local representative for this obstruction is the local functional
\[
\int_{\RR^2} J (Z') \, {\rm Tr}_{\fg} (A), 
\]
which depends solely on the degree $-1$ component $Z'$ of an element in~$\HH {\rm Vect}^{1,0} (\CC)$.
\end{prp}

Before we give the very short proof, we want to orient the reader familiar with the perturbative construction of 2-dimensional TFTs.
As a broad comment, AKSZ $\sigma$-models like the Poisson $\sigma$-model or topological $B$-model can be treated as variants of this topological BF theory:
in essence, one replaces the Lie algebra $\fg$ above by a bundle of $L_\infty$ algebras over a smooth manifold.
(To do this carefully, one uses Gelfand--Kazhdan geometry or $L_\infty$ spaces or Fedosov resolutions,
which are variations on the same theme.)
When one does this, one finds that an anomaly may emerge.
For concreteness' sake, we focus on the case of a topological $B$-model, where the target is a complex manifold~$X$.
In that case, the full anomaly to quantizing on an arbitrary oriented surface $S$ is $c_1(S) \otimes c_1(X)$, 
where $c_1$ denotes the first Chern class of a manifold.
To quantize one must fix a trivialization of that class (if it exists),
and so, to quantize on all surfaces, the manifold $X$ must be Calabi--Yau.
In \autocite{LiLi} this anomaly is computed using a gauge-fixing condition that exists on any genus $g$ Riemann surface. 

In our setting, the factor $\tr_\fg(A)$ plays the role of $c_1(X)$, 
as it depends on the target (encoded here by the gauge algebra $\fg$).
The Jacobian factor $J(Z')$ is the analog of $c_1(S)$,
as it is local on the source;
this factor arises because trivializing the action of holomorphic vector fields is needed to extend the theory to arbitrary Riemann surfaces. 

We wish to emphasize that our argument has a different flavor than that of~\autocite{LiLi}.

\begin{proof}
The proof is similar to the sorts of anomaly calculations we have already encountered in this section. 
The anomaly is represented by a two-vertex wheel with trivalent vertices, whose outer edges are labeled by a vector field $Z'$, which is degree~$-1$ in the dg Lie algebra $\HH {\rm Vect}^{1,0}(\CC)$, and the gauge field~$A$. 
The degree 0 component $Z$ of an input vector field $Z$ contributes nothing to the anomaly, 
as can be seen by spelling out the integral explicitly.

Let the input vector field have the form $Z' = f(z,\zbar) \frac{\partial}{\partial z}$.
As a function of the gauge field $A$, the anomaly is given by the integral
\[
\lim_{L \to 0} \lim_{\epsilon \to 0} \int_{\CC_z \times \CC_w} f(z,\zbar) {\rm Tr}_{\fg}(A(w, \Bar{w})) P_{\epsilon < L} (z,\zbar ;w, \Bar{w}) K_{\epsilon} (z,\zbar ; w , \Bar{w}) \d z .
\]
Using the formulas 
\begin{align*}
P_{\epsilon < L} (z,\zbar ;w, \Bar{w}) & = \int_{T=\epsilon}^L \frac{1}{4 \pi t} \frac{\zbar - \Bar{w}}{4 t} e^{-|z - w|^2 / 4t} \d T, \\
K_{\epsilon} (z,\zbar ; w , \Bar{w}) & = \frac{1}{4 \pi \epsilon} \frac{\zbar - \Bar{w}}{4 \epsilon} e^{-|z-w|^2 / 4 \epsilon} (\d \zbar - \d \wbar) ,
\end{align*}
one computes that the local representative for the anomaly is proportional to
\[
\left(\int_{\CC} \frac{\partial}{\partial z} f(z, \Bar{z}) {\rm Tr}_{\fg}(A) \right) \left(\lim_{\epsilon \to 0} \int_{t = \epsilon}^L \frac{\epsilon\, \d t}{(\epsilon + t)^2}\right) ,
\]
which is a nonzero multiple of $\int J(Z') {\rm Tr}_{\fg}(A)$, as desired.
\end{proof} 

\printbibliography
\end{document}